\documentclass[aps,nofootinbib,superscriptaddress, showpacs,preprintnumbers,nofootinbibt,twocolumn,prd]{revtex4-2}
\usepackage{eurosym}
\usepackage{graphicx}
\usepackage{amssymb}
\usepackage{amsmath}

\def\cL{\mathcal{L}}
\def\be{\begin{equation}}
\def\ee{\end{equation}}
\def\bea{\begin{eqnarray}}
\def\eea{\end{eqnarray}}

\begin{document}

\title{Cosmological implications of the Weyl geometric gravity theory}
\author{Tiberiu Harko}
\email{tiberiu.harko@aira.astro.ro}
\affiliation{Department of Theoretical Physics, National Institute of Physics
	and Nuclear Engineering (IFIN-HH), Bucharest, 077125 Romania,}
\affiliation{Department of Physics, Babes-Bolyai University, 1 Kogalniceanu Street,
	400084 Cluj-Napoca, Romania,}
\affiliation{Astronomical Observatory, 19 Ciresilor Street, 400487
	Cluj-Napoca, Romania,}
\author{Shahab Shahidi}
\email{s.shahidi@du.ac.ir}
\affiliation{School of Physics, Damghan University, Damghan, 41167-36716, Iran}

\begin{abstract}
We consider the cosmological implications of the Weyl geometric gravity theory. The basic action of the model is obtained from the simplest conformally invariant gravitational action, constructed, in Weyl geometry, from the square of the Weyl scalar, the strength of the Weyl vector, and a matter term, respectively. The total action is linearized in the Weyl scalar by introducing an auxiliary scalar field. To maintain the conformal invariance of the action the trace condition is imposed on the matter energy-momentum tensor, thus making the matter sector of the action conformally invariant. The field equations are derived by varying the action with respect to the metric tensor, the Weyl vector field, and the scalar field, respectively. We investigate the cosmological implications of the theory, and we obtain first the cosmological evolution equations for a flat, homogeneous and isotropic geometry, described by Friedmann-Lemaitre-Robertson-Walker metric,  which generalize the Friedmann equations of standard general relativity. In this context we consider  two cosmological models, corresponding to the vacuum state, and to the presence of matter described by a linear barotropic equation of state. In both cases we perform a detailed comparison of the predictions of the theory with the cosmological observational data, and with the standard $\Lambda$ CDM model. By assuming that the presence of the Weyl geometric effects induce small perturbations in the homogeneous and isotropic  cosmological background, and that the anisotropy parameter is small, the equations of the cosmological perturbations due to the presence of the Weyl geometric effects  are derived. The time evolution of the metric and matter perturbations are explicitly obtained. Therefore, if Weyl geometric effects are present, the Universe would acquire some anisotropic characteristics, and its geometry will deviate from the standard FLRW one.
\end{abstract}

\date{\today}
\maketitle

\affiliation{Department of Theoretical Physics, National Institute of Physics
and Nuclear Engineering (IFIN-HH), Bucharest, 077125 Romania,}
\affiliation{Astronomical Observatory, 19 Ciresilor Street, 400487
Cluj-Napoca, Romania,}
\affiliation{Department of Physics, Babes-Bolyai University, 1 Kogalniceanu Street,
400084 Cluj-Napoca, Romania,}
\affiliation{School of Physics, Sun Yat-Sen University,  510275 Guangzhou, People's
Republic of China.}

\affiliation{School of Physics, Damghan University, Damghan, 41167-36716,
Iran}


\section{Introduction}

Einstein's general relativity is a physical theory essentially based on the geometry of Riemann \cite{Riem}. The gravitational field equations, as proposed by Einstein \cite{Ein1, Ein2}, and Hilbert \cite{Hilb} provide a description of the gravitational field in which the contractions of the curvature tensor determine the gravitational interaction of massive objects. The gravitational force is thus not due to the microscopic exchange of particles (quanta of the fields), as is the case in electrodynamics, but it is an intrinsic property of space-time structure. General relativity has found extensive applications in astrophysics and cosmology, and it has been extensively tested at various scales. At the level of the Solar System, general relativity gives an excellent description of the of the gravitational dynamics, and fully explains the perihelion precession of Mercury, the bending of light by the Sun, or the Shapiro time delay effect \cite{Will1, Will2}. The experimental detection of the gravitational waves \cite{GW1, GW2} has also brilliantly confirmed the predictions of general relativity, and it has opened a new window for the understanding of the intricate physical processes taking black hole - black hole, or black hole - neutron star merging processes \cite{GW2}. One of the most intriguing predictions of general relativity is related to the existence of black holes, whose existence in the static spherically case was predicted by Schwarzschild \cite{Sch}, while the rotating black hole solution was obtained by Kerr \cite{Kerr}.  The first images of the supermassive black hole M87* were presented recently by the Event Horizon Telescope collaboration \cite{A1,A2,A3}. The observations point towards a Kerr-like structure for the Sgr A* black hole. But still one cannot reject the possible existence of differences with respect to the predictions of standard general relativity.

The impressive success, of general relativity as a physical theory of gravity based on geometric concepts, influenced significantly not only physics, but also mathematics. The differential geometric approach to gravity pioneered by Einstein led to a deeper understanding of the structure of Riemannian geometry, and opened the ways for its generalization. The possibility of solving physical problems by using geometrical methods suggested to study geometries that go beyond Riemann geometry, and general relativity. Historically, the first proposal to generalize Riemann geometry,  was due to Weyl \cite{W1,W2}. Weyl was motivated in this generalization by the intention of solving one of the most important problems of theoretical physics, namely,  the unification of the gravitational and electromagnetic forces. For an extensive account of the historical aspects of Weyl geometry, its evolution, and its physical and mathematical implications see \cite{W3}. A few years later after Weyl's work another important generalization of Riemann geometry was introduced  by Cartan \cite{Car1,Car2,Car3,Car4}, based on the concept of torsion. Geometries with torsion represent the mathematical foundations of the Einstein-Cartan theories of gravity \cite{Hehl1}. In these types of theories torsion describes the effects on the space-time geometry arising from the rotation of compact objects.

In building his geometry, Weyl adopted  two fundamental principles. The first is the possibility of the variation of the length of a vector during its parallel transport. The variation of the length is described by a geometric quantity called nonmetricity $Q_{\lambda \mu \nu}$.  Consequently, in Weyl geometry, the covariant derivative of the metric tensor $g_{\mu \nu}$ satisfies the condition $\tilde{\nabla}_\lambda g_{\mu \nu}=Q_{\lambda \mu \nu}$, also called the nonmetricity condition. Secondly, Weyl postulated that the laws of nature must be conformally invariant. In the first formulation of Weyl's geometry for the nonmetricity a particular form was adopted, namely,  $Q_{\lambda \mu \nu}=A_\lambda g_{\mu \nu}$, where $A_\mu$ is the Weyl vector. Weyl also proposed the interpretation of the vector $A_\mu$ as the electromagnetic field potential. However, the physical interpretation of Weyl's geometry was strongly disapproved  by Einstein. Due to this criticism, as well and as a result of the general development of theoretical physics, for a long time Weyl's unified field theory was not accepted as a possible physical approach to the unification of the electromagnetic and gravitational interactions. But Weyl's geometry has many beautiful characteristics, and it opens the way for the full implementation  of the conformal invariance of physical laws. Weyl's geometry is also at the origin of the gauge theory \cite{W4,W5}, which has become the fundamental theoretical tool in particle physics. Thus, in the past century, Weyl's geometry did continue to inspire the work of both physicists and mathematicians.

One of the interesting attempts for the reconsideration of Weyl's theory from a physical point of view was due to  Dirac \cite{Di1,Di2}, who introduced a real scalar field $\phi$ of weight $w(\phi)=-1$ in the basic theory. As for the gravitational Lagrangian Dirac proposed the expression
 $
 L=-\phi ^2R+kD^{\mu}\phi D_{\mu}\phi +c\phi ^4+W_{\mu \nu}W^{\mu \nu}/4,
 $
 where $R$ is the Ricci scalar, $c$ and $k=6$ are constants, and  $W_{\mu \nu}$ is the electromagnetic type tensor constructed from the Weyl vector. The Dirac Lagrangian has the important property of conformal invariance. A slightly modified version of the Dirac model was investigated, from a  cosmological perspective, in \cite{Ro}. The cosmological implications of a Weyl-Dirac type Lagrangian of the form
 $
 L=W^{\lambda \rho}W_{\lambda \rho}-\phi ^2R+\sigma \phi ^2w^{\lambda}w_{\lambda}+2\sigma \phi w^{\lambda}\phi _{,\lambda}+
 (\sigma +6)\phi _{,\rho}\phi_{,\lambda }g^{\rho \lambda}+2\Lambda \phi ^4+L_m,
 $
 where $\sigma$ and $\Lambda$ are constants, were investigated in \cite{Isr}. In the cosmological model based on the action introduced in \cite{Isr},  the creation of particles at the beginning of the Universe is determined by the Dirac gauge field.  During the late time evolutionary phases of the Universe,  the Dirac gauge field represents the source of the dark energy triggering the de Sitter exponential cosmic expansion of the Universe.

 Weyl's geometry was also generalized to include torsion. The corresponding geometry is called the Weyl-Cartan geometry, and it has also been extensively studied from both physical and mathematical points of view \cite{C1c,C2c,C3c,C4c,C5c,C6c,C7c,C8c, C9c}. For a detailed review of the geometry and of the cosmological  applications of the Riemann-Cartan and Weyl-Cartan space-times  see \cite{R}. The role of the torsion in the Weyl-Dirac theory was investigated  in \cite{Is1, Is2,Is3}. By including the torsion tensor in the theory one can also obtain a conformally invariant massive electrodynamic theory, generalizing classical electromagnetism.

 Another interesting  mathematical approach with important physical applications was initiated  by Weitzenb\"{o}ck \cite{Weitz}, who introduced the geometrical structure  known today as the Weitzenb\"{o}ck space. The Weitzenb\"{o}ck spaces have the basic properties $\nabla _{\mu }g_{\sigma \lambda }= 0$, $T^{\mu }_{\sigma \lambda }\neq 0$, and $R^{\mu }_{\nu \sigma \lambda }=0$, respectively, where $T^{\mu }_{\sigma \lambda }$ is the torsion tensor. Weitzenb\"{o}ck spaces reduce  to a Euclidean manifold when $T^{\mu }_{\sigma \lambda }= 0$. Moreover,  the torsion tensor $T^{\mu }_{\sigma \lambda }$ has distinct values in the different points of the manifold. Since the Riemann curvature tensor vanishes identically,  the Weitzenb\"{o}ck  spaces have the property of distant parallelism, also known as teleparallelism, or absolute parallelism. Einstein proposed a unified teleparallel theory of electromagnetism and gravitation based on the Weitzenb\"{o}ck geometry \cite{Ein}. Cosmological applications of the Wey-Cartan-Weitzenb\"{o}ck geometry were considered in \cite{WCWa,WCW1a}.

  A teleparallel  formulation of the gravitational field can be obtained by replacing the metric tensor $g_{\mu \nu}$  with a set of tetrad vectors $e^i_{\mu }$. Thus,  the torsion tensor, constructed from the tetrad fields, describes completely gravity, with the curvature replaced by the torsion. This approach represents the so-called teleparallel equivalent of General Relativity (TEGR), which was initially developed in \cite{TE1a,TE2a,TE3a}. The corresponding theory is  also known as the $f(\mathbb{T})$ gravity theory, with $\mathbb{T}$ denoting the torsion scalar. In the $f(\mathbb{T})$ type theories the equations describing the gravitational field are of  second order, while in other modified gravity theories, like, for example, the $f(R)$ gravity theory, the field equations in the metric formalism are of fourth order \cite{bookHLa}. See \cite{book1a} for a review of the TEGR type theories.  $f(\mathbb{T})$ theories can also explain geometrically the recent accelerated expansion of the Universe, without the need of introducing in the field equations a dark energy term, or the  cosmological constant \cite{T1a,T2a,T3a,T4a,T5a,T6a,T7a,T8a,T9a,T10a,T11a,T12a}.

 The determinations by the Planck satellite of the fluctuations in the temperature distribution of the Cosmic Microwave Background Radiation \cite{Ca1,Ca2}, as well as
the investigations of the light curves of the Type Ia supernovae \cite{Ca3}, have provided compelling evidence that the present day Universe is in a phase of rapid cosmological expansion. Moreover, these observational findings have also conclusively proven that only around 5\% of the matter-energy content of the Universe consists of baryonic matter, with 95\%  being represented by two other components, called generically dark matter, and dark energy, respectively. In order to interpret theoretically the cosmological observational  data, the $\Lambda$CDM ($\Lambda$ Cold Dark Matter) model was introduced, which is essentially based  on the reintroduction in the standard gravitational field equations of the cosmological constant $\lambda$,  first proposed in 1917 by Einstein \cite{L} to obtain a static cosmological model of the Universe. Later on, Einstein rejected the possibility of the existence of $\Lambda$. The $\Lambda$CDM model gives a very good description of the observational data, especially at low redshifts, and thus it is considered the standard cosmological paradigm of the present times. However, in its basic formulation the $\Lambda$CDM model is confronted with an important objection related to its theoretical foundation. Presently no satisfactory explanation of the nature (physical or geometrical) of the cosmological constant does exist, and thus the theoretical basis of the $\Lambda$CDM model are at least uncertain.

Thus, it is justifiable to suppose that to obtain a realistic physical picture of the Universe, fully consistent with the cosmological observations, one should extend standard general relativity. It may be possible that the Einstein field equations, describing very well the gravitational dynamics in the Solar System, are only a first order approximation of a more general gravity theory. Many modified gravity theories have been proposed, and for a detailed reviews see \cite{Od1,Od2,Od3}. Generalizations   of Einstein's relativity in the presence of geometry-matter couplings have been introduced in \cite{Ca4,Ca5, Ca6}. For a unified approach to modified gravity theories see \cite{B1,B2}.

The $\Lambda$CDM model is also confronted  with several other problems, which are mostly the results of the increase in accuracy of the cosmological observations. An important cosmological challenge is the difference between the expansion rate of the Universe as determined from the Cosmic microwave Background Radiation satellite observations, and the numerical values obtained from the local (low redshift) determinations \cite{O1}. This discrepancy is called as the Hubble tension \cite{O2,O3,O4}. The Hubble constant (H0) as measured by the Planck satellite, has the value of $66.93 \pm 0.62$ km/ s/ Mpc \cite{O5}, while the value of $73.24\pm 1.74$ km/ s/ Mpc
\cite{O1} is obtained by the SH0ES collaboration. The differences between these two values is more than 3$\sigma$ \cite{O6}. The Hubble tension, if indeed it exists,  points strongly towards the need of considering new gravitational theories, and extending, or even fully replacing, the $\Lambda$CDM model.

An important avenue for the  extension of standard cosmology, and for obtaining explanations of the present-day observations, is related to the reexamination of Weyl's theory as a possible description of the gravitational interaction.  A fundamental idea, first introduced by Weyl, is the consideration of the conformal invariance of the physical laws. It turns out that the fundamental equations of elementary particle physics are conformally invariant, but Einstein's general relativity is not. Hence, in order to make all the equations describing elementary particle interactions consistent, it is necessary to reformulate  general relativity as a conformally invariant field theory. Theories of gravity, satisfying the requirement of the conformal invariance, as well as conformally invariant theories of elementary particle physics  were proposed and investigated in detail in \cite{P1,P1a,P2,P3,P4,G1a,G2}.

A fully conformally invariant theory of gravity can be obtained  from the action
$
S=-\alpha _g\int{C_{\lambda \mu \nu \kappa}C^{\lambda \mu \nu \kappa}\sqrt{-g}d^4x}
=-2\alpha _g\int{\left(R_{\mu \nu}R^{\mu \nu}-R^2/3\right)\sqrt{-g}d^4x},
$
where $C_{\lambda \mu \nu \kappa}$ is the (conformally invariant) Weyl tensor, and $\alpha _g$ is a constant. This theory was introduced, and extensively investigated, in  \cite{Ma0,Ma1,Ma2,Ma3,Ma4,Ma5}. The theory can also provide an explanation for the unusual motion of the massive particles orbiting around the galactic centers, which is usually explained by assuming the existence of a mysterious, and yet undetected, component of the Universe, called dark matter.

 Weyl geometry represents the mathematical basis of the $f(Q)$ modified gravity theory \cite{Q1, Q2, Q4}, and of its generalizations \cite{Q10, Q17, Q20, Q23, Q24}.  In the $f(Q)$ theory, the basic geometric parameter, fully describing the gravitational interaction, is the non-metricity $Q$. The action of the $f(Q)$ theory is obtained as
 $
 S=\int{f(Q)\sqrt{-g}d^4x},
 $
 where $f(Q)$ denotes an arbitrary analytical function of $Q$. An extension of the $f(Q)$ theory, based on the action $S=\int{f(Q,T)\sqrt{-g}d^4x}$ where $T$ is the trace of the matter energy-momentum tensor, was introduced in \cite{Q17}. The $f(Q)$ theory, as well as its extensions, were intensively investigated from both theoretical and observational points of view \cite{Laz,C0a,Pinto, C1a,C2a,C3a,C4a, Mandal}.

A novel, and very interesting perspective on Weyl gravity, and of its applications in elementary particle physics, cosmology and astrophysics,  was proposed, and extensively developed,  in \cite{Gh1, Gh2,Gh3,Gh4,Gh5,Gh6,Gh7,Gh8, Gh9, Gh10, Gh11,Gh12,Gh13}. This approach to gravity heavily relies on concepts from elementary particle physics, and in the following we will call it the Weyl geometric gravity theory (sometimes it is also called quadratic Weyl gravity \cite{Gh12}). The starting point in this approach to Weyl gravity is to linearize, in the quadratic Weyl action \cite{W1,W2}, the square of the Weyl scalar  $\tilde{R}^2$ via the introduction of an auxiliary scalar field $\phi$. As a result, quadratic Weyl gravity can be reformulated  as a gravitational theory linear in the curvature scalar. This linearization has important physical implications.  In the curvature linearized Weyl action one can introduce a spontaneous breaking of the $D(1)$ symmetry as a result of the presence of a Stueckelberg type mechanism, having a geometric origin. Consequently, the Weyl gauge vector field acquires a mass, originating from the spin-zero mode of the $\tilde{R}^ 2$ term. The Stueckelberg mechanism is implemented via the replacement of the scalar field $\phi$ with a constant value (its vacuum expectation value), with $\phi\rightarrow <\phi>$.  Hence, through this mechanism the Weyl vector field  becomes massive, and in this way the dynamical scalar field $\phi$ is absorbed in the mass of the vector field. Consequently, the scalar field is eliminated from the initial scalar-vector-tensor theory. After removing $\phi$, the Einstein-Proca action is obtained from the initial Weyl action. Therefore, a vector-tensor theory is reobtained, which is similar to the initial Weyl geometric gravity theory.

The pathway to gravity via the linearization procedure of the quadratic Weyl action also provides some insights on the Planck scale, and on the cosmological constant. We have already mentioned that the Einstein-Proca action emerges from the initial Weyl action in the broken phase. Consequently, all mass scales, including the Planck scale, as well as the cosmological constant $\Lambda$, have a geometric origin  \cite{Gh8}. The Higgs field, which plays an essential role in the standard model of elementary particles, also originates from geometry, and it is created through the fusion of Weyl bosons in the very early Universe, during the reheating phase.

The physical and cosmological properties of the Weyl geometric gravity have been extensively investigated recently. The conformally invariant coupling between geometry and matter  was considered in \cite{C1, C2, C3}. The Palatini formulation of the theory in the presence of conformally invariant matter-geometry couplings was studied in  \cite{C2}. The Palatini formulation of the quadratic Weyl gravity $\tilde{R}^2+R_{\mu\nu }^2$ was also analyzed in \cite{Gh5}, by assuming that the metric and the Weyl connection are independent quantities. In the Palatini  approach all the mass scales do appear as a purely geometric effects. Moreover,  a spontaneous breaking of the gauge scale symmetry can be implemented in the theory. An inflationary scenario can also be constructed, and the tensor-to-scalar ratio is predicted as $0.007 \leq r \leq 0.01$, at 95 \% CL, and $N =60$ efolds. The investigation of the inflation in the Weyl geometric gravity in its metric and Palatini versions was initiated in \cite{Gh6}. Since the two versions of the Weyl geometric gravity have different non-metricities, both determined by the Weyl gauge field, the physical predictions of the two formalisms are different. Black hole solutions in Weyl geometric gravity were obtained, by using both numerical and analytical methods, in \cite{C4}. The possibility that dark matter is a Weyl geometric effect was considered in \cite{C5}. Stellar type objects in Weyl geometric gravity were studied in \cite{C6}. The spin Hall effect for light was generalized to the case of Weyl geometry in \cite{C7}. The thermodynamical properties of the Weyl geometric black holes were analyzed extensively in \cite{C8}.  The behavior of the galactic rotation curves in Weyl geometric gravity were investigated in \cite{C9}.

It is the goal of the present paper to consider isotropic and anisotropic cosmological models in the simplest model of the Weyl geometric gravity. Our starting point is a gravitational action consisting of the sum of the square $\tilde{R}^2$ of the Weyl scalar, and of the field strength $F_{\mu \nu}^2$ of the Weyl vector. The action can be linearized in the Weyl scalar, by introducing an auxiliary scalar field $\phi$. Then the conformally invariant Weyl action can be reformulated as an effective scalar-vector-tensor theory in Riemann geometry, with the action containing effective couplings between the scalar field and the Ricci scalar, and the Weyl vector field. These terms are conformally invariant by construction.  Moreover, a matter term is also added to the total action in a conformally invariant way. The field equations corresponding to this action are obtained in the metric formalism, by varying the action respect to the metric tensor, Weyl vector and the scalar field.

An important question in conformally invariant gravitational actions is how to implement the conformal invariance of the matter terms. In this work, we will achieve the requirement of the conformal invariance of the matter action by imposing the trace condition on the effective matter action $\mathcal{L}_m$, constructed with the help of the ordinary matter action $L_m$, and the square of the Weyl vector $\omega ^2$, so that $\mathcal{L}_m=\mathcal{L}_m\left(L_m,\omega ^2\right)$. Once this condition is satisfied, the corresponding gravitational field equations, and their solutions,  are conformally invariant.

After obtaining the gravitational field equations of the Weyl geometric gravity, and the consistency condition in the homogeneous and isotropic Friedmann-Lemaitre-Robertson-Walker cosmological framework (the generalized Friedmann equations), we consider a number of specific cosmological models. Thus, we investigate first a vacuum cosmological model, in which the matter Lagrangian and the baryonic matter density are assumed to vanish. The cosmological equations are solved numerically, and the Hubble function of the model is compared with a small observational dataset of observational values of the Hubble function, as well as with the predictions of the standard $\Lambda$CDM paradigm. The behaviors of the deceleration parameter, of the Weyl vector, and of the scalar field are also obtained.   

The effect of the matter is considered in the next step of our investigation. The generalized Friedmann equations are again solved numerically, and the comparison with the observations and $\Lambda$CDM model is performed in detail. The variations with respect to the redshift of the matter energy density, deceleration parameter, Weyl vector and scalar field are also considered. 

  Finally, we investigate the anisotropic properties of the Weyl geometric cosmological models. We consider the simplest extension of the isotropic and homogeneous FLRW geometry, namely Bianchi type I spacetimes. We write down the gravitational equations describing the Weyl geometric evolution in a Bianchi type I geometry in an effective form, by introducing the equivalent dark energy and dark matter terms, which also incorporate the contributions from Weyl geometry. By assuming that the differences between  the scale factor of the isotropic FLRW model and the scale factors of the Bianchi type I spacetime are small, it follows that the deviations from isotropy are also small, and therefore we can assume that they represent just a small perturbation of the homogeneous and isotropic FLRW background metric. We derive the basic equations satisfied  by the perturbations of the metric, and of the effective (geometric) energy densities and pressures. It turns out that the cosmological evolution of the perturbed geometric and physical quantities is determined by the Hubble function of the background isotropic model. The perturbation equations are solved numerically, for the case of the Bianchi type I geometry, and the behaviors of the relevant physical and geometrical quantities, like the anisotropy parameter, deceleration parameter or energy density and pressure perturbations are obtained.

As a possible test of the anisotropic Weyl geometric cosmological model we have considered the behavior of the quadrupole $Q_2$ of the CMBR as function of the deviations $\delta _1$ and $\delta _3$ from the isotropic and homogeneous  FLRW geometry. Since the dynamical characteristics  of the background FLRW geometry in Weyl geometric gravity are known, one could use the obtained expression of $Q_2$ to put tight constraints on the parameters of the anisotropic Weyl geometric cosmology model. For the isotropic case we have obtained the constraints on the model parameters from the study of the luminosity distance by using the observational data from the type Ia supernovae.

The present paper is organized as follows. The action and the field equations of the Weyl geometric gravity theory are introduced in Section~\ref{sect1}, where the conformally invariant construction of the matter action is also discussed. The cosmological evolution of the isotropic and homogeneous Weyl geometric gravity models is discussed in Section~\ref{sect2}. The generalized Friedmann equations are obtained, and two cosmological models, corresponding to a vacuum Universe, and a matter filled one, are explored in detail. A comparison with the observational data for the Hubble function, and with the $\Lambda$CDM model is also performed. Bianchi type I homogeneous and anisotropic cosmological models are considered in Section~\ref{sect3}, with the use of a perturbative approach. The behavior of the various geometrical and cosmological quantities describing the model are also considered in detail. Finally, we discuss and conclude our results in Section~\ref{sect4}.  

\section{Action and field equations of Weyl geometric gravity}\label{sect1}

in the present Section we write down first the action of Weyl geometric gravity, quadratic in the Weyl scalar. Secondly, we briefly introduce the basic concepts of Weyl geometry to be used in the sequel. The problem of the conformally invariant coupling of ordinary matter is also considered, and the trace condition is obtained in a general form. Then, the field equations of the theory are obtained by varying the Weyl type action with respect to the metric.

\subsection{From the gravitational Lagrangian to Weyl geometry}

The  most general gravitational Lagrangian density that is invariant under a gauged Weyl symmetry, and defined in Weyl geometry, is given by \cite{Gh6,Gh7,Gh8}
\bea\label{L1}
\cL_1=\sqrt{g}\,\,\Big\{\frac{1}{4!\,\xi^2} \tilde R^2-\frac{1}{\eta^2}\, \tilde C_{\mu\nu\rho\sigma}^2
-\frac{1}{4}\,F_{\mu\nu}^2\Big\}, 0<\xi,\,\eta <1, \nonumber\\
\eea
were $F_{\mu\nu}$ is the field strength of the Weyl gauge field, and $\tilde C_{\mu\nu\rho\sigma}$ is the Weyl tensor.

A local Weyl symmetry is defined as the invariance of the action under the set of transformations
\bea\label{WG}
\hat g_{\mu\nu}&=&\Sigma^q (x)g_{\mu\nu}, \quad
\sqrt{-\hat g}=\Sigma^{2\,q}(x)\sqrt{-g},
\eea
\bea\label{WGa}
\hat\phi&=&\Sigma^{-q/2}(x)\phi,\quad \quad
\hat\psi=\Sigma^{-3 q/4}(x) \psi,
\eea
where $\Sigma (x)$ is an arbitrary positive definite function of the coordinates, $q$ is a constant called the Weyl charge of the metric, and $\phi$ and $\psi$ are bosonic and fermionic fields, respectively.

The gauged Weyl symmetry (or Weyl gauge symmetry for short)
is defined as the invariance of the action under the transformations (\ref{WG}), and the transformation 
\bea\label{WGg}
\hat\omega_\mu=\omega_\mu-\frac{1}{\alpha}\,\partial_\mu\ln\Sigma,
\eea
where $\omega _\mu$  is the associated Weyl gauge field, while $\alpha$ denotes the Weyl gauge coupling constant.

In the following,  by Weyl geometry we consider a geometry that is  invariant under the transformations (\ref{WG}), (\ref{WGa}), and (\ref{WGg}), respectively. Also, for simplicity, and without loss of generality, we will consider only the case $q=1$. The Weyl geometry is non-metric, and the covariant divergence of the metric tensor satisfies the condition
\bea\label{wge}
\tilde\nabla_\mu g_{\alpha\beta}=-\alpha\,q\, \omega_\mu \,g_{\alpha\beta},
\eea
where
\bea
\tilde\nabla_\mu g_{\alpha\beta}\equiv \partial_\mu g_{\alpha\beta}
-\tilde\Gamma_{\alpha\mu}^\rho g_{\rho\beta}
-\tilde\Gamma_{\beta\mu}^\rho g_{\rho\alpha},
\eea
and $\tilde\Gamma_{\beta\mu}^\rho$ are the coefficients of the Weyl connection.

The Weyl connection $\tilde\Gamma$ can be found  by direct calculation from Eq.~(\ref{wge}), and it is given by
\bea\label{tGammap}
\tilde \Gamma_{\mu\nu}^\lambda&=&\Gamma_{\mu\nu}^\lambda\Big\vert_{\partial_\mu\rightarrow \partial_\mu +\alpha\, \omega_\mu}\nonumber\\
&=&
\Gamma_{\mu\nu}^\lambda+\frac{\alpha}{2}\, \,\Big(\delta_\mu^\lambda\,\, \omega_\nu +\delta_\nu^\lambda\,\, \omega_\mu
- g_{\mu\nu} \,\omega^\lambda\Big),
\eea
where the Levi-Civita connection $\Gamma_{\mu\nu}^\lambda$ is defined by
\bea
\Gamma_{\mu\nu}^\lambda=\frac{1}{2}
\,g^{\lambda\alpha}(\partial_\mu g_{\alpha\nu} +\partial_\nu g_{\alpha\mu}-
\partial_\alpha g_{\mu\nu}).
\eea

The scalar curvature $\tilde R$ of the Weyl geometry and the square of the Weyl tensor $\tilde C_{\mu\nu\rho\sigma}^2$ can be obtained as \cite{Gh10}
\bea\label{tildeR}
\tilde R=R-3\,\alpha \,\nabla_\mu\omega^\mu-\frac{3}{2} \alpha ^2\,\omega_\mu \,\omega^\mu, \\
\tilde C_{\mu\nu\rho\sigma}^2=C_{\mu\nu\rho\sigma}^2+\frac{3}{2} \,\alpha^2 \,F_{\mu\nu}^2.
\eea
In the following the quantities without tilde represent the Riemannian geometric counterparts of the geometric quantities defined in Weyl geometry.

Since
$\tilde \Gamma_{\mu\nu}^\lambda=\tilde \Gamma_{\nu\mu}^\lambda$, then
$F_{\mu\nu}=
\tilde\nabla_\mu \omega_\nu-\tilde\nabla_\nu \omega_\mu
=\partial_\mu\omega_\nu-\partial_\nu \omega_\mu$, identical to $F_{\mu\nu}$ as defined in the  Riemann (or flat) space-time.

We proceed now to the scalar-vector-tensor representation of the Weylian Lagrangian density (\ref{L1}).  To obtain this representation we perform the substitution \cite{Gh6,Gh7,Gh8}
\be
\tilde R^2\rightarrow - 2 \phi^2 \tilde R-\phi^4,
\ee
in the gravitational Lagrangian (\ref{L1}), where $\phi$ is an auxiliary scalar field. Moreover, we neglect in the gravitational Lagrangian the term containing the Weyl tensor. This is equivalent in taking the limit $\eta \rightarrow \infty$ in our model. Hence, the action of the Weyl geometric gravity theory can be written as
\begin{widetext}
\begin{align}\label{act}
	S=\int d^4x\sqrt{-g}\left[\frac{1}{12\xi^2}\phi^2\left(R-3\alpha\nabla_\mu \omega ^\mu-6\alpha^2\omega ^2-\frac{1}{2}\phi^2\right)-\frac14F^2+\mathcal{L}_m\left(L_m,\omega ^2,\psi\right)\right],
\end{align}
\end{widetext}
where $\psi$ is an external field, $L_m$ is the
ordinary (baryonic) matter Lagrangian, and $\mathcal{L}_m$ is the effective
matter Lagrangian of the Weyl geometric theory, respectively. The effective matter Lagrangian  $\mathcal{L}_m$ must contain minimal or non-minimal couplings between the
matter Lagrangian and the Weyl geometric quantities in order to assure the conformal invariance of the theory.
Moreover, we have denoted $\omega ^2\equiv \omega _{\mu}\omega ^\mu$, and $F^2\equiv
F_{\mu\nu}F^{\mu\nu}$.

\subsection{Conformal coupling of matter}

It is important to note at this moment that it is not necessary for the matter part $\mathcal{L}_m$ of the action
(\ref{act}) to be gauge invariant with respect to
the conformal transformations. However, the variation of $\mathcal{L}_m$ must have this important
invariance property \cite{Bera,Berb,Berc,Berd}. By varying the matter action in
Eq.~(\ref{act}) we obtain
\begin{eqnarray}
\delta S_{m}&=&-\frac{1}{2}\int T^{\mu \nu (tot)}\delta g_{\mu \nu }\sqrt{-g}%
d^{4}x+\int G^{\mu }\delta \omega  _{\mu }\sqrt{-g}d^{4}x  \notag \\
&&+\int \frac{\delta \mathcal{L}_{m}\left(L_m,\omega ^2,\psi\right)}{\delta \psi }%
\delta \psi ,
\end{eqnarray}
where by $T^{(tot)}_{\mu\nu}$ we denote the effective total energy-momentum tensor,
defined according to
\begin{equation}
T^{(tot)}_{\mu\nu}=-\frac{2}{\sqrt{-g}}\frac{\delta\left[ \sqrt{-g}
\mathcal{L}_m\left(L_m,\omega  ^2,\psi\right)\right]}{\delta g^{\mu\nu}},
\end{equation}
and
\begin{equation}
G^{\mu}\left(L_m,\omega ^2,\omega ^{\mu},\psi\right)=\frac{\delta \mathcal{L}%
_m\left(L_m,\omega  ^2,\psi\right)}{\delta \omega  ^{\mu}},
\end{equation}
denotes the Weyl current \cite{Bera,Berb,Berc,Berd}. In the following we assume
that $\delta \mathcal{L}_m/\delta \psi =0$.

With the use of the mathematical results \cite{Bera,Berb,Berc,Berd}
\begin{equation}
\delta g_{\mu \nu }=-\frac{2\delta \Omega }{\Omega ^{3}}\tilde{g}_{\mu \nu
}=-2\frac{\delta \Omega }{\Omega }g_{\mu \nu },
\end{equation}
\begin{eqnarray}
\delta \omega  _{\mu }&=&-\frac{4}{\alpha }\delta \frac{\partial _{\mu }\Omega }{%
\Omega }=-\frac{4}{\alpha }\delta \left( \partial _{\mu }\ln \Omega \right)
=-\frac{4}{\alpha }\partial _{\mu }\left( \delta \ln \Omega \right)  \notag
\\
&=&-\frac{4}{\alpha }\partial _{\mu }\left( \frac{\delta \Omega }{\Omega }%
\right)=-\frac{4}{\alpha }\nabla _{\mu }\left( \frac{\delta \Omega }{\Omega }%
\right) ,
\end{eqnarray}
we obtain for the variation of the matter action the relation
\begin{eqnarray}  \label{Cond}
\delta S_{m}&=&\int T^{\mu \nu (tot)}g_{\mu \nu }\frac{\delta \Omega }{%
\Omega }\sqrt{-g}d^{4}x  \notag \\
&&+\frac{4}{\alpha }\int G^{\mu }\nabla _{\mu }\left( \frac{\delta \Omega }{%
\Omega }\right) \sqrt{-g}d^{4}x=0.
\end{eqnarray}

After performing a partial integration of the second term in the above equation, and with the use of the Gauss
theorem, from Eq.~(\ref{Cond}) we obtain the important consistency (trace) condition that must be satisfied by the matter terms of the Weyl geometric gravity theory
\begin{align}  \label{tr1}
T^{(tot)}=-\frac{8}{\alpha}\nabla_\mu\left[\omega^\mu \frac{\partial\mathcal{L%
}_m\left(L_m, \omega ^2, \psi\right)}{\partial \omega^2}\right],
\end{align}
where $T^{(tot)}=g^{\mu\nu}T^{(tot)}_{\mu\nu}$ is the trace of the effective
energy-momentum tensor, obtained with the help of the effective matter
Lagrangian $\mathcal{L}_m$.

From Eq.~\eqref{tr1} it immediately follows that when $\mathcal{L}_m=L_m$, the constraint (\ref{tr1}) gives the
 usual trace condition $T^{(m)}=0$, where by $T^{(m)}$ we have denoted  the trace of the ordinary
matter energy-momentum tensor. This leads to the important result that the only conformally
invariant form of ordinary matter must have a traceless energy-momentum tensor, thus corresponding to a radiation type fluid.

\subsection{The gravitational field equations of Weyl geometric gravity}

In this paper, we will assume that the total (effective) matter Lagrangian is given by
\begin{align}
\mathcal{L}_m=L_m+\beta (-\omega ^2)^n,
\end{align}
where $L_m$ is the baryonic matter Lagrangian, and $\beta $ and $n$ are constants.

Before considering the cosmological implications of the model, let us transform the action to a form which is easier to handle in cosmology. If one defines the new set of quantities $\left(\Phi,A_\mu,\beta _2\right)$, given by 
\begin{align}
	\Phi=\phi^2,\quad A_\mu=\alpha\omega_\mu,
	\quad \beta_2=12\beta\xi^2\alpha^{-2n},
\end{align}
one can write the action as
\begin{widetext}
\begin{align}\label{Lf}
		S=\frac{1}{12\xi^2}\int d^4x\sqrt{-g}\left[\Phi\left(R-3\nabla_\mu A^\mu-6A^2-\frac{1}{2}\Phi\right)-\frac{3\xi^2}{\alpha^2}F^2+\mathcal{L}_m\left(L_m,A^2,\psi\right)\right],
\end{align}
\end{widetext}
with the matter Lagrangian given by
\begin{align}
	\mathcal{L}_m=12\xi^2 L_m+\beta_2(-A^2)^n.
\end{align}

Also the trace constraint equation is simplified as
\begin{align}  \label{tr}
	T^{(tot)}=-8\nabla_\mu\left(A^\mu \frac{\partial\mathcal{L%
		}_m\left(L_m, A ^2, \psi\right)}{\partial A^2}\right),
\end{align}

Varying the  action (\ref{Lf}) with respect to $g_{\mu\nu}$, $\Phi$ and $A_{\mu}$
gives the full set of the field equations of the Weyl geometric gravity as 
\bea
&&	\Phi G_{\mu\nu}-\frac32 \Phi \left(A_\mu A_\nu-\frac12 A^2 g_{\mu\nu}\right)+\frac14 \Phi g_{\mu\nu}+\Box\Phi g_{\mu\nu}\nonumber\\
&&-\nabla_\mu\nabla_\nu\Phi-\frac{6\xi^2}{\alpha^2} \left(F_\mu^{~\alpha}F_{\nu\alpha}-\frac14 F^2 g_{\mu\nu}\right)
	-6\xi^2 T_{\mu\nu}\nonumber\\
&&-\beta_2 nX\left(A_\mu A_\nu-\frac{1}{2n} A^2 g_{\mu\nu}\right)\nonumber\\
&&+\frac32 \left(A_\nu\nabla_\mu\Phi+A_\mu\nabla_\nu\Phi-g_{\mu\nu}A^\alpha\nabla_\alpha\Phi\right)=0,
\eea
\begin{align}
	\frac{12\xi^2}{\alpha^2}\nabla_\alpha F^{\mu\alpha}+3\Phi A^\mu-3\nabla^\mu\Phi+2\beta_2 n X A^\mu=0,
\end{align}
\begin{align}\label{Sf}
	R-\frac32 A^2-3\nabla_\mu A^\mu-\Phi=0,
\end{align}
where we have defined $X\equiv (-A^2)^{n-1}$. 

The trace equation
can also be written as
\begin{align}
3\xi^2T&+\beta_2 (-A^2)^{n-4}\Bigg(2n(n-1)A^\mu\nabla A^2\nonumber\\
&-(n-2)(-A^2)^2 +2nA^2\nabla_\mu A^\mu\Bigg)=0.
\end{align}

Using the trace of the metric equation to eliminate $R$ in the scalar field equation (\ref{Sf}), one obtains
\begin{align}\label{box}
	\Box  \Phi-\nabla _\rho \left(\Phi A^\rho\right)-\frac{\beta_2}{3}(n-2)XA^2-2\xi^2 T=0,
\end{align}

\section{Cosmological implications of Weyl geometric gravity - the case of the FLRW geometry}\label{sect2}

In the present Section we consider the cosmological implications of the Weyl geometric gravity in the presence of baryonic matter coupled in a conformally invariant way to the Weyl geometric quantities in a homogeneous and isotropic Universe. As a first step in our study we obtain the generalized Friedmann equations of Weyl geometric gravity. The vacuum case and the cosmological evolution in the presence of matter are both considered. In each case we compare the predictions of the theory with the observational data.

\subsection{The generalized Friedmann equations}

We assume first that the geometry of the Universe can be described by a flat FLRW ansatz of the form
\begin{align}
ds^2=-dt^2+a ^2(t)\left(dx^2+dy^2+dz^2\right),
\end{align}
where $a $ is the scale factor. We also introduce the Hubble function (parameter) $H$ defined as $H=\dot{a}/a$. In the following by a dot we denote the derivative with respect to the time $t$. We will assume that the energy
momentum tensor has a perfect fluid form, and in the comoving frame it is given by
\begin{align}
T^\mu_{~\nu}={{\rm diag}}(-\rho,p,p,p),
\end{align}
where $\rho $ and $p$ denote the baryonic matter energy density and pressure, respectively. For the matter Lagrangian we adopt the expression  $L_m=-\rho$. With the above assumptions we can
write
\begin{align}
A_{\mu}=(A_0,\vec{0}),
\end{align}
where $A_0$ is a function of the cosmological time $t$ only.

The cosmological field equations of the Weyl geometric gravity theory can then be written as
\begin{align}
	\Phi(12H^2&-3A_0^2-\Phi)-24\xi^2\rho+6(2H+A_0)\dot\Phi \nonumber\\
&+2\beta_2(1-2n)A_0^2X=0,
\end{align}
\begin{align}
	2\Phi\dot{H}&+\frac32A_0^2\Phi+6\xi^2(\rho+p)-\left(H+3A_0\right)\dot\Phi+\ddot\Phi\nonumber\\
&+\beta_2 nA_0^2X=0,
\end{align}
\begin{align}
	\dot\Phi-A_0\Phi-\frac23\beta_2 nA_0X=0,
\end{align}
\begin{align}
	6\dot{H}+3\dot{A}_0-\Phi+\frac32(A_0+2H)(A_0+4H)=0,
\end{align}
and
\begin{align}
3\xi^2(3p&-\rho)\nonumber\\&=\beta_2 X\left[(n-2)A_0^2-6nA_0H+2n(1-2n)\dot{A}_0\right],
\end{align}
respectively. 

Now, combining the above set of equations, one can obtain the conservation
equation of the matter energy momentum tensor as
\begin{align}\label{conser}
\dot\rho+3H(\rho+p)=0,
\end{align}
which shows that the matter sector is conserved in this theory. This is however obvious from the action \eqref{act}, since there is no non-minimal coupling between the  ordinary baryonic matter, and the gravitational fields.

\subsection{The vacuum solution}

We will first investigate the cosmological evolution in Weyl geometric
gravity by neglecting the effect of the baryonic matter, that is, by assuming $\rho =p=0$, and $L_{m}=0$, respectively. In this case the cosmological evolution equations
become
\begin{align}\label{eqa1}
	2\Phi\dot{H}+\frac32A_0^2\Phi-(H+3A_0)\dot\Phi+\ddot\Phi+\beta_2 nA_0^2X=0,
\end{align}
\begin{align}\label{eqa2}
	\dot\Phi-A_0\Phi-\frac23\beta_2 nA_0X=0,
\end{align}
\begin{align}\label{eqa3}
	6\dot{H}+3\dot{A}_0-\Phi+\frac32(A_0+2H)(A_0+4H)=0,
\end{align}
and
\begin{align}\label{eqa4}
	\beta_2 X\left((n-2)A_0^2-6nA_0H+2n(1-2n)\dot{A}_0\right)=0,
\end{align}
respectively. As can be seen from the above set of equations, we have more equations than variables. In the following we will solve numerically the equations (\ref{eqa1})-(\ref{eqa3}). We will then substitute the solution into equation \eqref{eqa4} to check whether the extra equation is also satisfied.

We define now the following set of dimensionless variables
\begin{align}
	\tau &=H_0t,\quad \quad H=H_0 h,  \notag \\
	A_0&=H_0\bar{A}_0,\quad
	\beta_2 = \gamma H_0^{4-2n},\quad \Phi=H_0^2\bar\Phi,
\end{align}
where $H_0$ is the present day value of the Hubble function. Moreover, instead of the time coordinate, we introduce the redshift coordinate $z$, defined as $1+z=1/a$.

Hence,  one
obtains the cosmological field equations of the Weyl geometric gravity in the vacuum in a dimensionless form as
	\begin{align}		5(2n\gamma\bar{X}&+3\bar\Phi)\bar{A}_0^2-4n(2n-1)(1+z)\gamma\bar{X}h\bar{A}_0^\prime\nonumber\\&+6(1+z)\left(h^\prime\bar\Phi^\prime-h\bar\Phi(\bar{A}_0^\prime+2h^\prime)\right)\nonumber\\&+18(1+z)\bar{A}_0h\bar\Phi^\prime=0,
	\end{align}
	\begin{align}
		-\frac23\bar\Phi+\bar{A}_0^2+6\bar{A}_0h+8h^2-2(1+z)h(\bar{A}_0^\prime+2h^\prime)=0,
	\end{align}
\begin{align}
	3(1+z)h\bar\Phi^\prime+3\bar{A}_0\bar\Phi-2\gamma n\bar{X}\bar{A}_0=0,
\end{align}
and
	\begin{align}
		\gamma\bar{X}\left(12n\bar{A}_0h-4n(2n-1)(1+z)h\bar{A}_0^\prime-(n-2)\bar{A}_0^2\right)=0,
	\end{align}
respectively. Here, we have denoted $\bar{X}=\bar{A}_0^{2n-2}$. Also, using the relation $h(z=0)=1$, one can obtain the value of the parameter $\gamma$ in terms of the present day value of the scalar and vector fields as
	\begin{align}
		\gamma=\frac{\bar{A}_0(0)^{1-2n}(\bar\Phi(0)^2-3(2+\bar{A}_0(0))^2\bar\Phi(0))}{2(\bar{A}_0(0)+4n)}.
	\end{align}

In order to find the best fit value of the parameters $H_0$, $\bar\Phi_0\equiv\bar{\Phi}(0)$, $\bar{A}_0(0)$ and $n$, we use the Likelihood analysis using the observational data on the Hubble parameter in the redshift range $z\in(0.07,2.36)$  tabulated in \cite{hubble}.  

In the case of independent data points, the likelihood function can be defined as
\begin{align}
	L=L_0e^{-\chi^2/2},
\end{align}
where $L_0$ is the normalization constant, and the quantity $\chi^2$ is defined as
\begin{align}
	\chi^2=\sum_i\left(\frac{O_i-T_i}{\sigma_i}\right)^2.
\end{align}
Here $i$ counts the data points, $O_i$ are the observational value, $T_i$ are the theoretical values, and $\sigma_i$ are the errors associated with the $i$th data obtained from observations.

By maximizing the likelihood function, the best fit values of the parameters $\bar\Phi_0\equiv\bar{\Phi}(0)$, $\bar{A}_0(0)$, $n$ and $H_0$ at $1\sigma$ confidence level, can be obtained as
\begin{figure}
	\includegraphics[scale=0.4]{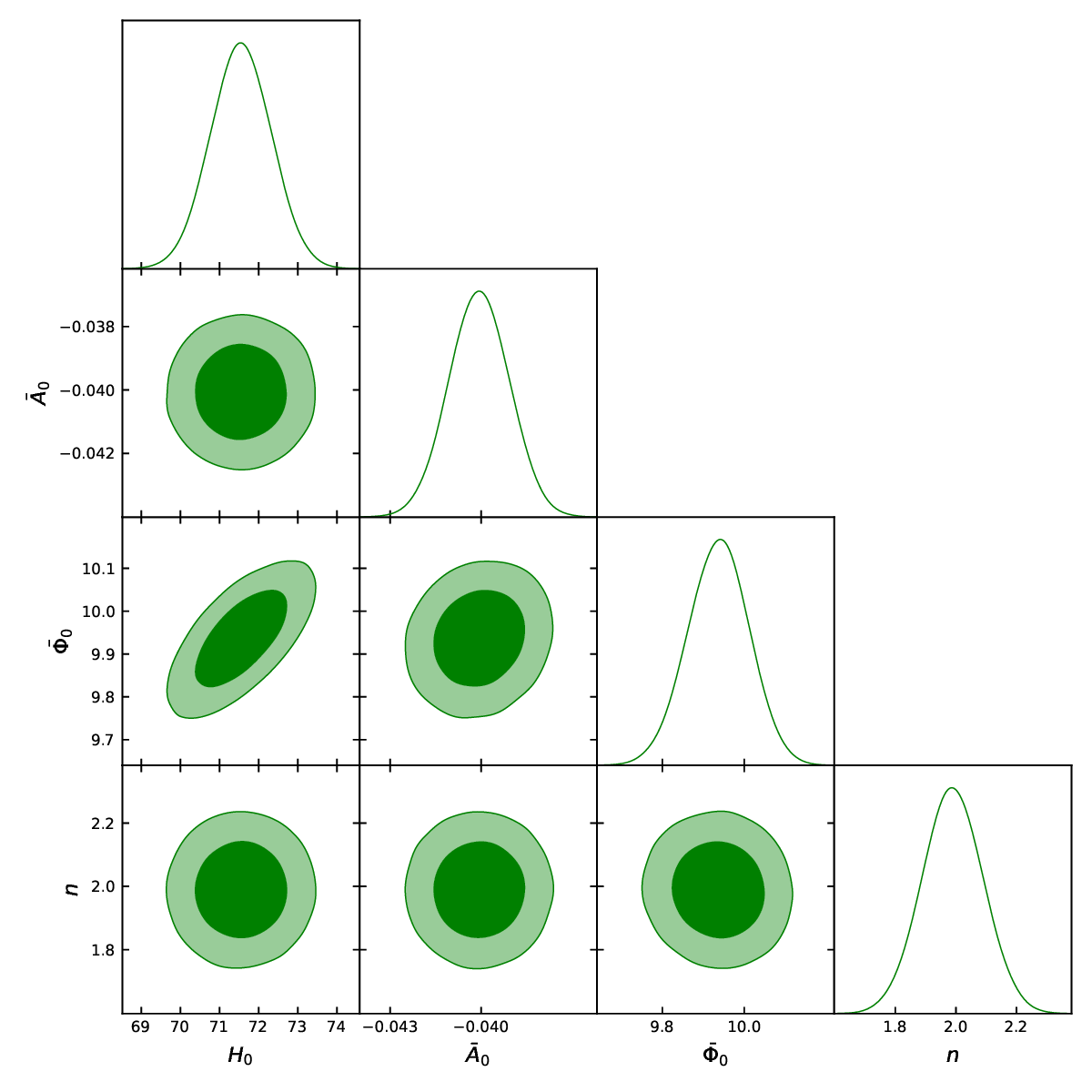}
	\caption{The corner plot for the values of the parameters $H_0$, $\bar\Phi_0\equiv\bar{\Phi}(0)$ and $\bar{A}_0(0)$ with their $1\sigma$ and $2\sigma$ confidence levels for the vacuum solution of the FLRW Weyl geometric gravity model. \label{cornervac}}
\end{figure}
\begin{align}\label{bestvac}
	H_0&=71.547^{+0.779}_{-0.772},\nonumber\\
\bar{A}_0(0)&=-0.040^{+0.001}_{-0.001},\nonumber\\
\bar\Phi_0&=9.938^{+0.073}_{-0.076},\nonumber\\
	n&=1.989^{+0.101}_{-0.100}.
\end{align}

The corner plot for the values of the parameters $H_0$, $\bar\Phi_0$ and $\bar{A}_0(0)$ with their $1\sigma$ and $2\sigma$ confidence levels is shown in Fig.~\ref{cornervac}.

\begin{figure*}
	\includegraphics[scale=0.5]{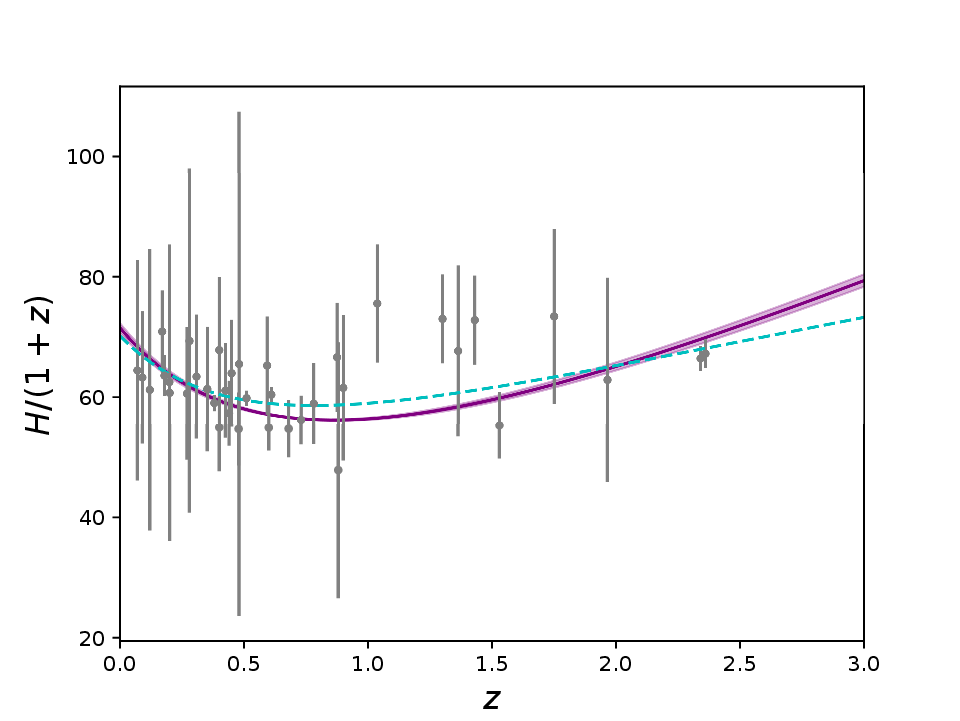}\includegraphics[scale=0.5]{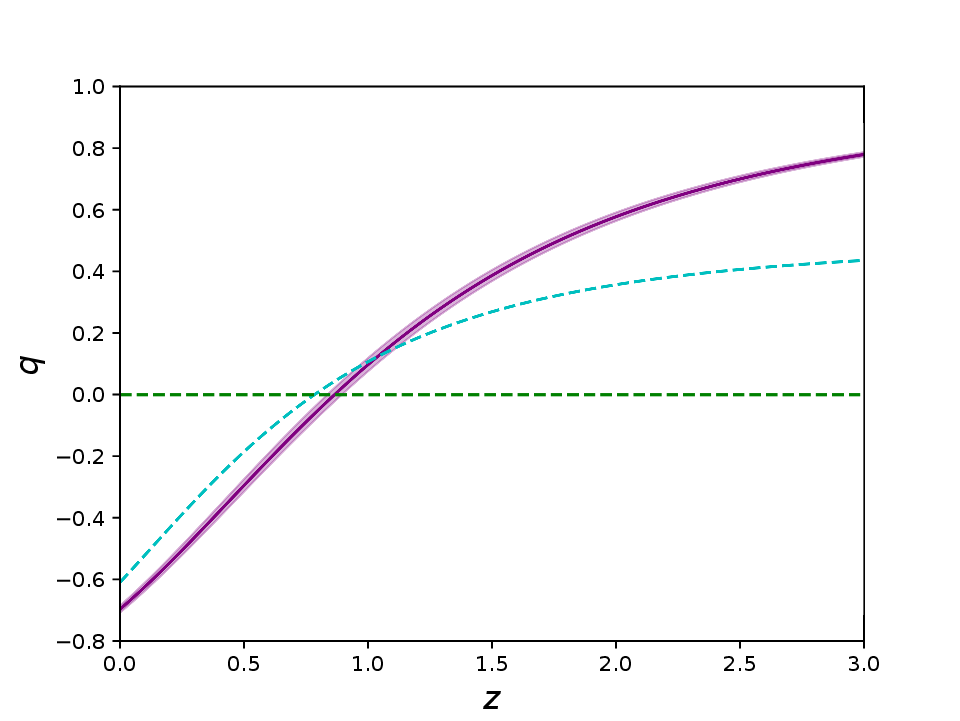}
	\caption{\label{fig1vac} The behavior of the rescaled Hubble parameter $H/(1+z)$ (left panel) and of the deceleration parameter $q$ (right panel) as a function of the redshift for the vacuum FLRW Weyl geometric gravity model for the best fit values of the parameters as given by Eqs.~(\ref{bestvac}). The shaded area denotes the $1\sigma$ error. The dashed line represents the $\Lambda$CDM model.}
\end{figure*}

\begin{figure*}
	\includegraphics[scale=0.5]{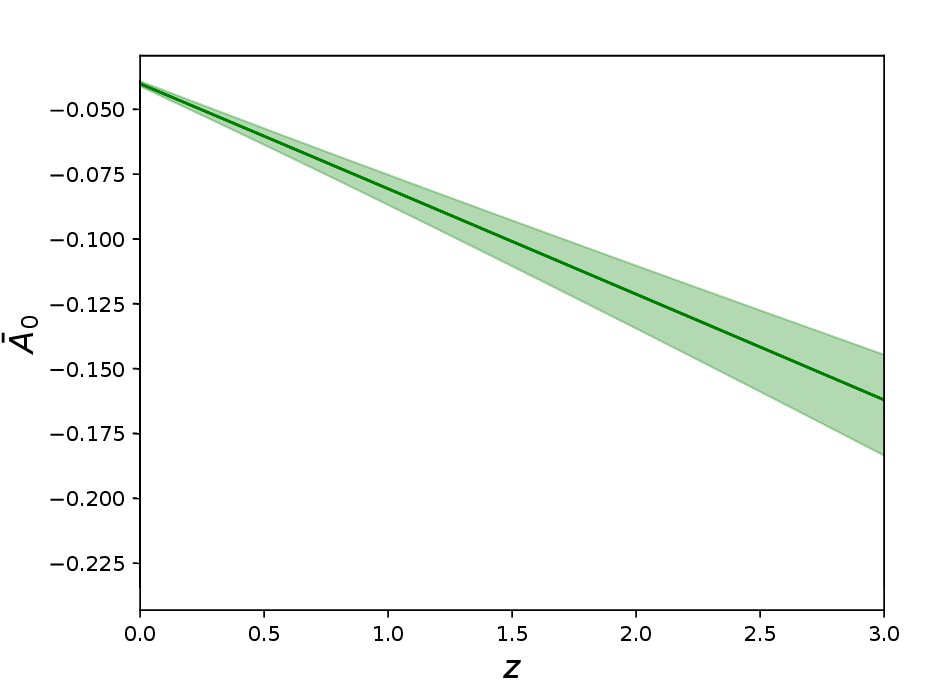}\includegraphics[scale=0.5]{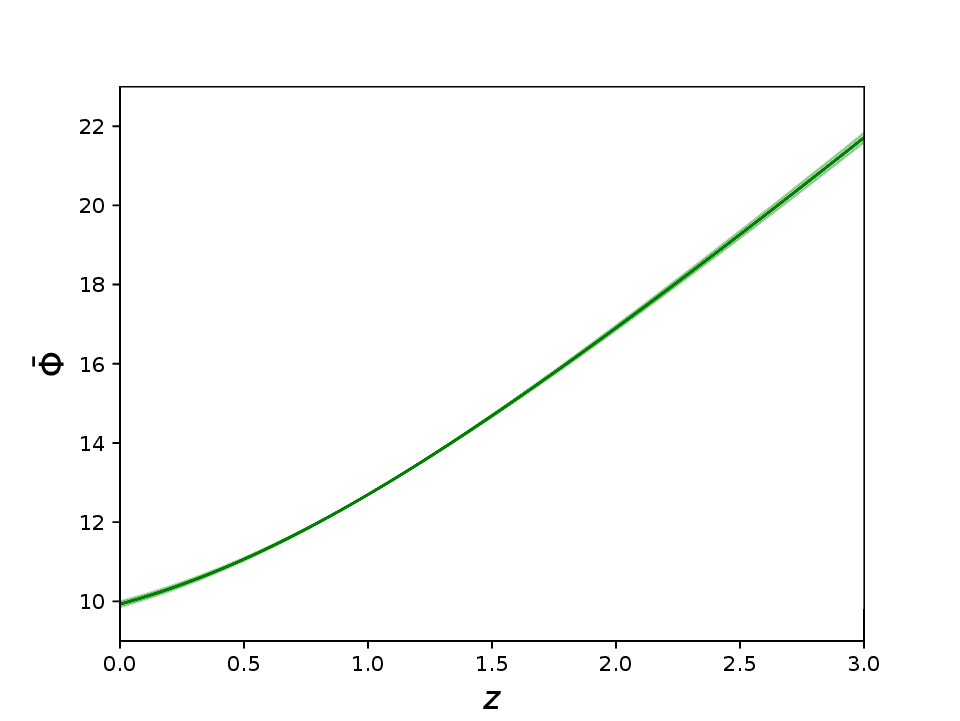}
	\caption{\label{fig2vac} The behavior of $\bar{A}_0$ (left panel) and of $\bar\Phi$ (right panel) as a function of the redshift in the vacuum FLRW Weyl geometric gravity theory for the best fit values of the parameters as given by Eqs.~(\ref{bestvac}).  The shaded area denotes the $1\sigma$ error.}
\end{figure*}

The redshift evolution of the Hubble function and of the deceleration parameter $q$ are represented, for this model, in Fig.~\ref{fig1vac}. Also, we have depicted the behavior of the cosmological fields $\bar\Phi$ and $\bar{A}_0$ in Fig.~\ref{fig2vac}.

\subsection{Cosmological evolution in the presence of matter}

We assume now that the equation of state of the baryonic matter fields is given by $%
p =\omega_m \rho $, with a varying equation of state parameter. We let the observations to determine $\omega_m $.  From the matter conservation equation, one can obtain
\begin{align}
	\omega_m=-1-\frac{\dot\rho}{3H\rho}.
\end{align}
After defining a set of dimensionless parameters as
\begin{align}
\tau &=H_0t,\quad H=H_0 h,\quad \bar\rho=\frac{2\xi^2}{H_0^2}\rho,  \notag \\
A_0&=H_0\bar{A}_0,\quad
 \beta_2 = \gamma H_0^{4-2n},\quad \Phi=H_0^2\bar\Phi,
\end{align}
and transforming to the redshift coordinate, defined again as $1+z=1/a$, one
obtains the cosmological field equations (generalized Friedmann equations) in the presence of baryonic matter  as
\bea
&&5(2n\gamma\bar{X}+3\bar\Phi)\bar{A}_0^2-4n(2n-1)(1+z)\gamma\bar{X}h\bar{A}_0^\prime \nonumber\\
&&+18(1+z)\bar{A}_0h\bar\Phi^\prime \nonumber\\
&&+6(1+z)\left[h^\prime\bar\Phi^\prime-h\bar\Phi(\bar{A}_0^\prime+2h^\prime)+\bar\rho^\prime\right]=0,
\eea
\begin{align}
	-\frac23\bar\Phi+\bar{A}_0^2+6\bar{A}_0h+8h^2-2(1+z)h(\bar{A}_0^\prime+2h^\prime)=0,
\end{align}
\bea
&&	\gamma\bar{X}\left(12n\bar{A}_0h-4n(2n-1)(1+z)h\bar{A}_0^\prime-(n-2)\bar{A}_0^2\right)\nonumber\\
&&-12\bar\rho+3(1+z)\bar\rho^\prime=0,
\eea
and
\begin{align}
	3(1+z)h\bar\Phi^\prime+3\bar{A}_0\bar\Phi-2\gamma n\bar{X}\bar{A}_0=0,
\end{align}
respectively, where we have defined $\bar{X}=\bar{A}_0^{2n-2}$. Noting that $h(z=0)=1$, one can obtain the value of the parameter $\gamma$ in terms of the present value of the fields as
\begin{align}
	\gamma=\frac{\bar{A}_0(0)^{1-2n}(\bar\Phi(0)^2-3(2+\bar{A}_0(0))^2\bar\Phi(0)+12\bar\rho(0))}{2(\bar{A}_0(0)+4n)}.
\end{align}

By maximizing the likelihood function, the best fit values of the parameters $\bar\Phi_0\equiv\bar{\Phi}(0)$, $\bar{A}_0(0)$, $\bar\rho_0$, $n$ and $H_0$ at $1\sigma$ confidence level, can be obtained as
\begin{figure*}
		\includegraphics[scale=0.5]{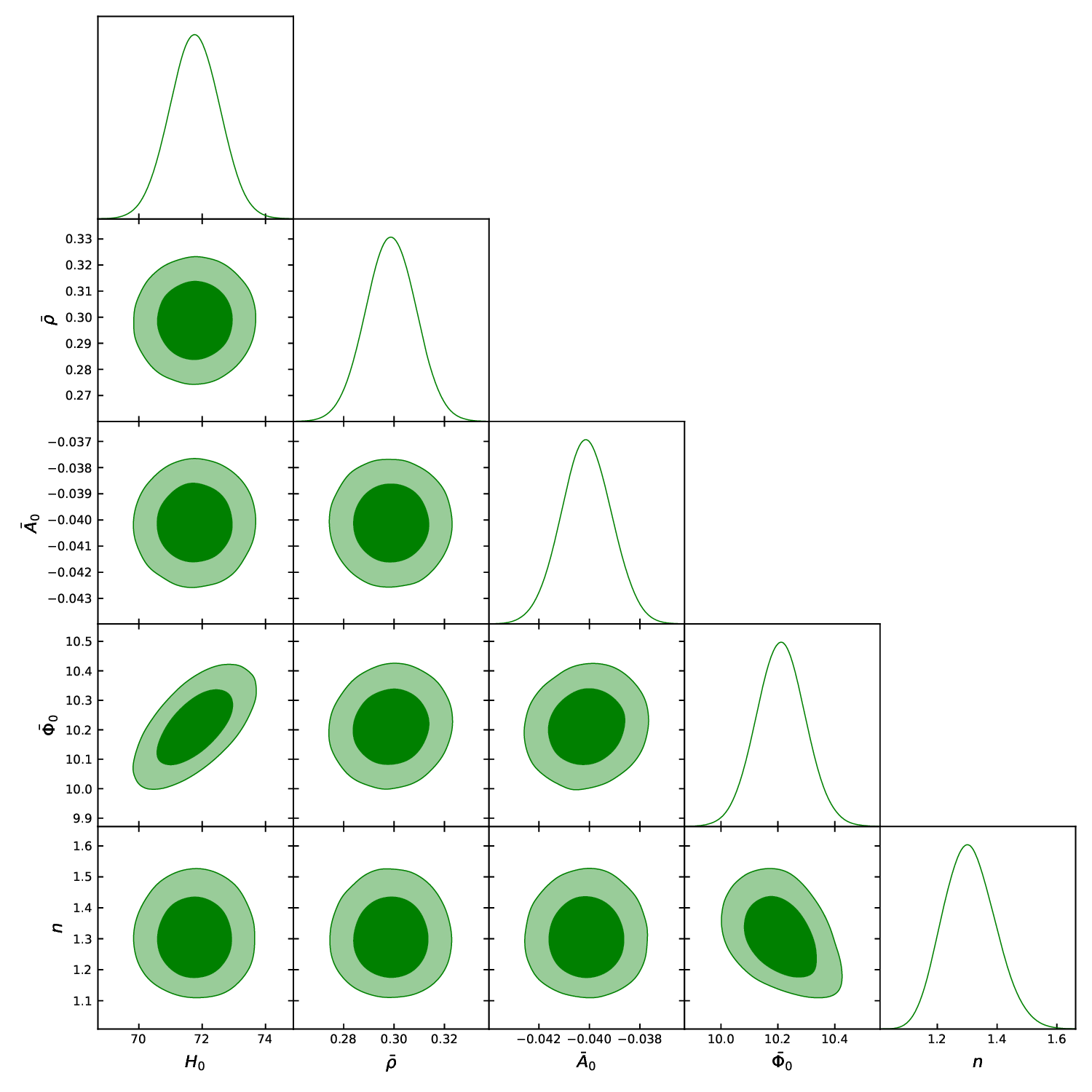}
		\caption{The corner plot for the values of the parameters $H_0$, $\bar\rho_0$,$\bar\Phi_0\equiv\bar{\Phi}(0)$ and $\bar{A}_0(0)$ with their $1\sigma$ and $2\sigma$ confidence levels for the FLRW Weyl geometric gravity model in the presence of matter. \label{corner}}
\end{figure*}

\begin{align}\label{best}
	\bar\rho_0&=0.299^{+0.001}_{-0.001},\nonumber\\
	H_0&=71.768^{+0.784}_{-0.787},\nonumber\\
	\bar{A}_0(0)&=-0.040^{+0.010}_{-0.010},\nonumber\\
	\bar{\Phi}(0)&=10.210^{+0.085}_{-0.085},\nonumber\\
	n&=1.305^{+0.090}_{-0.084}.
\end{align}

The corner plot for the values of the parameters $H_0$, $\bar\rho_0$, $\bar\Phi_0$, $\bar{A}_0(0)$ and $n$ with their $1\sigma$ and $2\sigma$ confidence levels is shown in Fig.~\ref{corner}.

\begin{figure*}
	\includegraphics[scale=0.5]{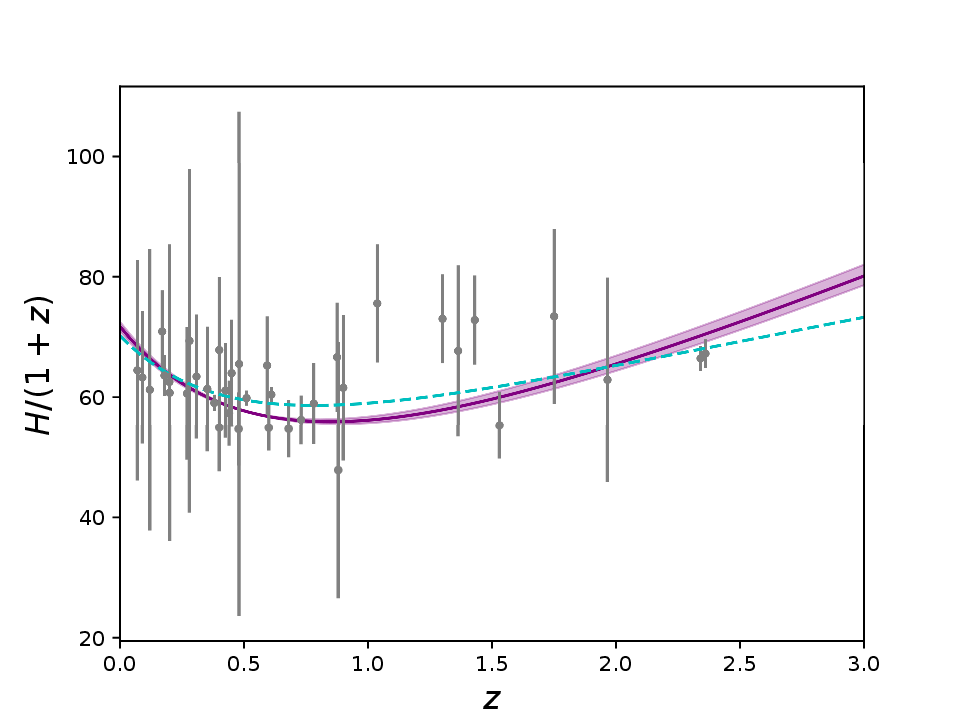}\includegraphics[scale=0.5]{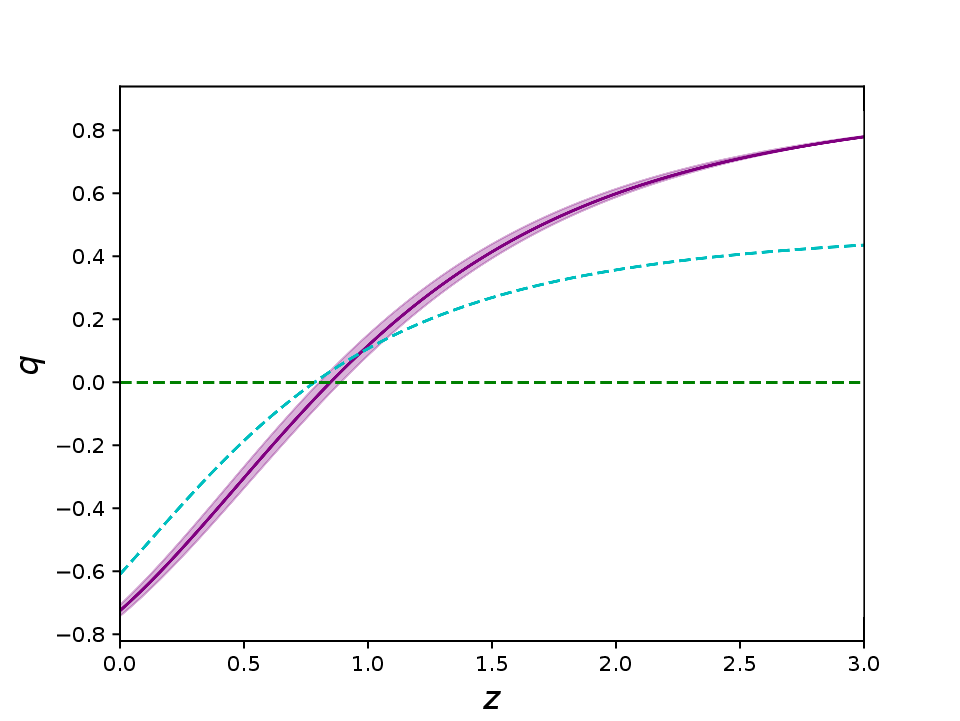}
	\caption{\label{fig1} The behavior of the rescaled Hubble parameter $H/(1+z)$ (left panel) and of the deceleration parameter $q$ (right panel) as a function of the redshift for the best fit values of the parameters as given by Eqs.~(\ref{best}) for the FLRW Weyl geometric gravity theory in the presence of matter. The shaded area denotes the $1\sigma$ error. The dashed line represents the $\Lambda$CDM model.}
\end{figure*}

\begin{figure*}
	\includegraphics[scale=0.5]{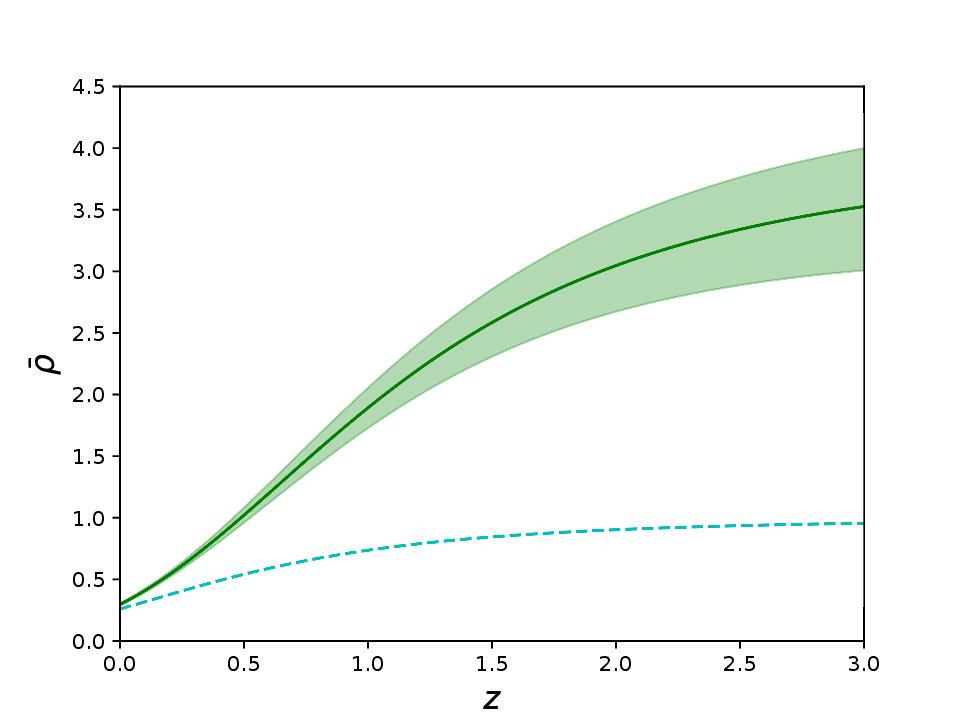}\includegraphics[scale=0.5]{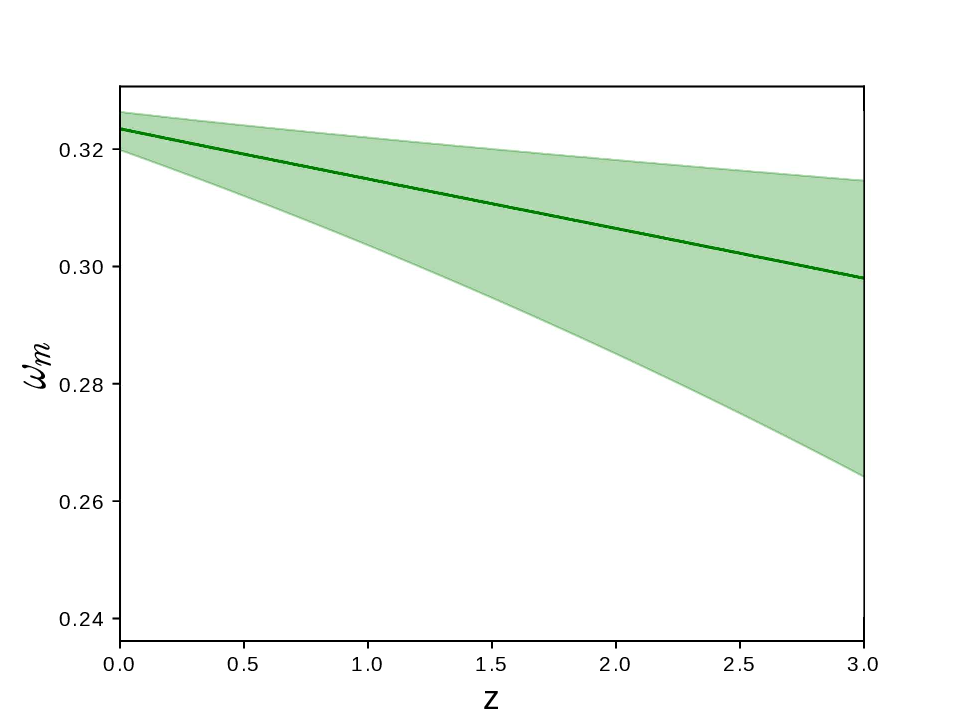}
	\caption{\label{fig2} The behavior of the matter density abundance $\bar\rho$ (left panel) and of the equation of state parameter $\omega$ (right panel) as a function of the redshift for the best fit values of the parameters as given by Eqs.~(\ref{best}) for the FLRW Weyl geometric gravity theory in the presence of matter.  The shaded area denotes the $1\sigma$ error. The dashed line represents the $\Lambda$CDM model.}
\end{figure*}

\begin{figure*}
	\includegraphics[scale=0.5]{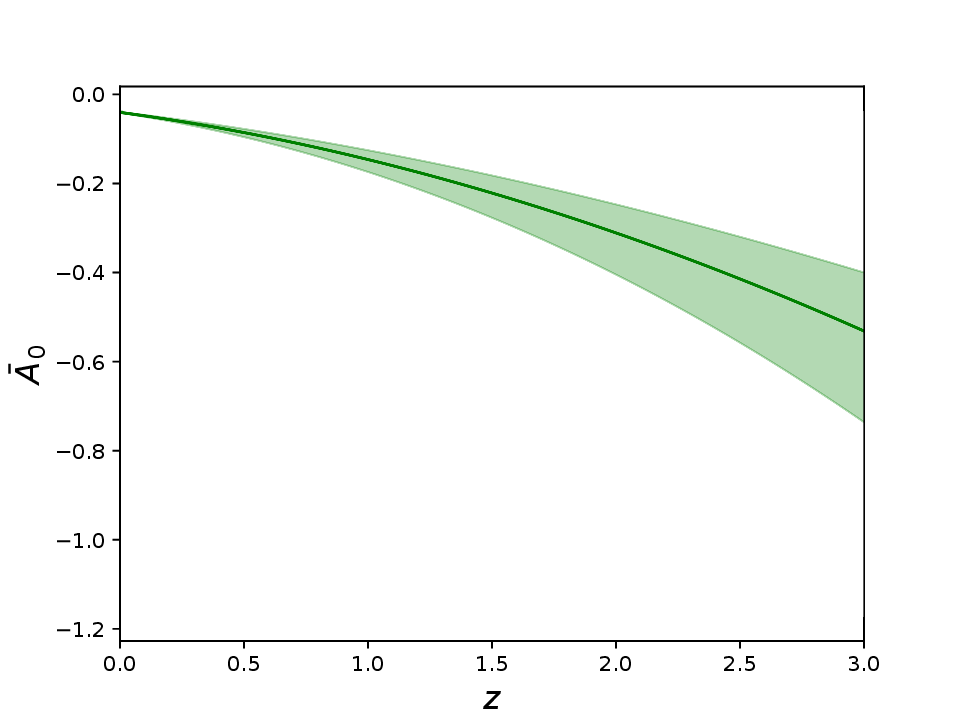}\includegraphics[scale=0.5]{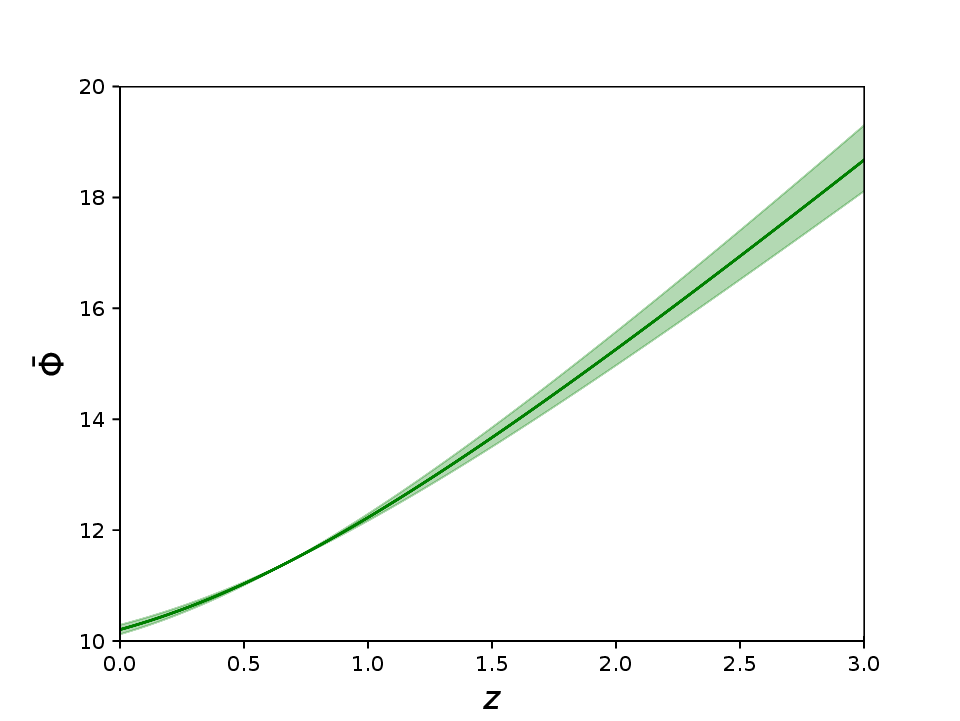}
	\caption{\label{fig3} The behavior of $\bar\Phi$ (left panel) and $\bar{A}_0$ (right panel) as a function of the redshift for the best fit values of the parameters as given by Eqs.~(\ref{best}) for the FLRW Weyl geometric gravity theory in the presence of matter.  The shaded area denotes the $1\sigma$ error.}
\end{figure*}

The redshift evolution of the Hubble function, of the deceleration parameter $q$, and of the matter density parameter $\Omega=\bar\rho/h^2$ are represented, for this model, in Figs.~\ref{fig1} and \ref{fig2}, respectively. Also, we have depicted the behavior of the cosmological fields $\bar\Phi$ and $\bar{A}_0$ in Fig.~\ref{fig3}. One can compute the value of the equation of state parameter $\omega_m$ as
\begin{align}
	\omega_m=0.3216^{+0.0001}_{-0.0001}.
\end{align}


\section{Cosmological implications of Weyl geometric gravity - the Bianchi type I geometry}\label{sect3}

In the present Section, we consider the cosmological applications Of the Weyl geometric gravity theory in an anisotropic cosmological setting. We begin our analysis by writing down the gravitational field equations corresponding to an anisotropic expansion of the Universe. We will consider only the case of flat Bianchi type I geometries, which are the simplest anisotropic generalizations of the flat FLRW geometry. As a first step in our analysis we analyze the general properties of the anisotropic Bianchi type I model in Weyl geometric gravity. In order to simplify our analysis, a perturbative approach is developed.

\subsection{Gravitational field equations in the Bianchi type I geometry}

In the following, we assume that on the large scale the Universe is homogeneous, and thus all the cosmological and physical parameters (metric, energy densities, pressures, Weyl vector, and scalar field) are functions of the cosmological time $t$ only.  In comoving Cartesian coordinates $x^{0}=t$, $x^{1}=x$, $%
x^{2}=y $, and $x^{3}=z $,  and in an anisotropic geometry, the components of the ordinary matter energy-momentum tensor take the general form
\begin{equation}\label{tenss}
T_{0}^{0}=-\varepsilon, \qquad  T_{1}^{1}=T_2^2=\Pi , \qquad T_{3}^{3}=\Psi ,
\end{equation}
where $\varepsilon $ denotes the total (effective) energy density of the cosmological fluid, which includes also the Weyl geometric contributions,  $\Pi =P_{x }=P_y$ denote the  effective pressure along the $x$ and $y$ directions, while  $\Psi =P_{z}$ is the pressure along the $z$
direction. All the effective pressure terms include also the geometric contributions coming from Weyl geometry. 

Since the energy-momentum tensor (\ref{tenss}) corresponds to an anisotropic effective cosmological fluid, the geometry must also be anisotropic on a large cosmological scale.  From our choice of the components of the energy-momentum tensor, as given by Eqs.~(\ref{tenss}), it follows that the effective  pressure $\Psi$ along the $z-$ axis could be different with respect to the effective pressures $\Pi$ along the  $x$ and $y$ axis.

The simplest geometry presenting the symmetry (\ref{tenss}) of the effective energy-momentum tensor in a homogeneous Universe is the flat Bianchi type I geometry, with the metric represented by
\begin{equation}
ds^{2}=-dt^{2}+a_{1}^{2}(t) dx^{2}+a_{2}^{2}(t) dy^{2}+a_{3}^{2}(t) dz^{2},
\label{7}
\end{equation}%
where $a_i $, $i=1,2,3$ are the three directional scale factors, which are generally different. The Einstein field equations
\begin{align}
	R_{\mu\nu}=\frac{1}{2\kappa^2}\left(T_{\mu\nu}-\frac12 Tg_{\mu\nu}\right),
\end{align}
 take in a Bianchi type I  geometry the form
\begin{equation}
3\dot{H}+H_{1}^{2}+H_{2}^{2}+H_{3}^{2}=-\frac{1}{4\kappa^2}\left( \varepsilon +\Psi
+2\Pi \right) ,  \label{8}
\end{equation}%
\be
\frac{1}{V}\frac{d}{dt}\left( VH_{i}\right)  =\frac{1}{4\kappa^2}\left(
\varepsilon -\Psi \right) , \qquad i=1,2 , \label{9}
\ee
and
\be\label{101}
\frac{1}{V}\frac{d}{dt}%
\left( VH_{3}\right) =\frac{1}{4\kappa^2}\left( \varepsilon +\Psi-2\Pi \right) ,
\ee
respectively,  where, in order to simplify the mathematical formalism,  we have introduced the following notations
\begin{equation}
V=a_{1}a_{2}a_{3}, \qquad H_{i}=\frac{\dot{a}_{i}}{a_{i}}, \qquad i=1,2,3,
\end{equation}
and
\begin{equation}
H=\frac{1}{3} \left(\sum_{i=1}^{3}H_{i}\right)=\frac{\dot{V}}{3V},
\end{equation}
respectively. 

$H_i$, $i=1,2,3$ are the three directional Hubble parameters, while by $H$ we have denoted  the mean Hubble parameter. The cosmological expansion parameter $\theta $ is related to the Hubble parameter by the relation $\theta =3H$. The conservation of the total energy-momentum tensor of the Weyl geometric gravity cosmological model gives the evolution equation of the effective energy density of the cosmic fluid as
\be
\dot{\varepsilon}+3\left(\varepsilon +\Pi\right)H+\left(\Psi -\Pi \right)H_3=0.
\ee

\subsection{General properties of the Bianchi type I models}

 By taking into account  the explicit expression of the anisotropic energy-momentum tensor as described by Eqs.~(\ref{tenss}), from Eqs.~(\ref{9}) it follows that we can take $a_1 =a_2 $ without any loss of generality. Hence $\theta$,  the expansion parameter of the Universe, is represented as
\be\label{th}
\theta =2\frac{\dot{a}_1}{a_1}+\frac{\dot{a}_3}{a_3}.
\ee
The shear scalar $\sigma $ of the Universe is obtained in the form
\be\label{sh}
\sigma =\frac{1}{\sqrt{3}}\left(\frac{\dot{a}_3}{a_3}-\frac{\dot{a}_1}{a_1}\right).
\ee

Eqs.~(\ref{th}) and (\ref{sh}) determine the relation between the directional Hubble parameters $H_1$ and $H_2$ and the observable cosmological parameters as
\be
H_1=H_2=\frac{\theta }{3}-\frac{1}{\sqrt{3}}\sigma,\qquad  H_3=\frac{\theta }{3}+\frac{2}{\sqrt{3}}\sigma,
\ee
respectively.

The late-time evolution of an anisotropic cosmological model can be obtained from the study of another important cosmological quantity, the  anisotropy
parameter $A$, defined according to
\begin{equation}
A=\frac{1}{3}\sum_{i=1}^{3}\left( \frac{H_{i}-H}{H}\right) ^{2}.
\end{equation}
If the anisotropy parameter vanishes, $A=0$, the considered cosmological model is isotropic.

The  differences of the effective pressures of the cosmological fluid are given, as functions of the observable quantities, by the general relation
\be
\Psi -\Pi =2\sqrt{3}\kappa^2(\theta \sigma+\dot\sigma).
\ee

By adding Eqs.~(\ref{9}) and (\ref{101})  we obtain
\be\label{41}
4\kappa^2\ddot{V}+\left(2\Pi+\Psi -3\varepsilon \right)V=0.
\ee
These expressions will be considered in the next Section.

\subsection{Perturbative approach to the Bianchi type I cosmological models}\label{sectn}

In the present Section, we consider a simple and elementary perturbative approach to the gravitational field equations (\ref{8})-(\ref{101}), by assuming that the anisotropic expansion due to the presence of the Weyl geometric effects represents a small perturbation of the background flat, and isotropic FLRW geometry. 

Hence, the cosmological properties of the background geometry are described by the isotropic scale factor $a $. Moreover, we suppose that the cosmological expansion rates along the $x$ and $y$ axes are the same, which implies the condition $a_1 =a_2 $. Furthermore, we assume that the perturbations along these two axes are also equal. Therefore, the scale factors of the perturbed Bianchi type I geometry can be represented as \cite{HL}
\be
a_i =a +\delta _i , \quad \delta _i\ll a,  \qquad i=1,2,3,
 \ee
 where $\delta _i $, $i=1,2,3$, are small corrections terms to the scale factor $a $ of the isotropic Universe, induced by the existence of the Weyl geometric effects. Our previous  assumption implies
 \be
 \delta _1 =\delta _2 .
 \ee

For the background isotropic FLRW type geometry the Hubble parameter is $H_0 =\dot{a}/a$. Hence,  for the directional Hubble
parameters $H_i$ of the Bianchi type I geometry we obtain
 \bea\label{Hi}
 H_i =\frac{\dot{a} +\dot{\delta}_i }{a +\delta _i }\approx H_0 \left[1-\frac{\delta _i }{a }\right]+\frac{\dot{\delta _i} }{a },i=1,2,3.\nonumber\\
  \eea
 The expression of the mean Hubble parameter $H$ of the perturbed Bianchi type I geometry in the presence of Weyl geometric effects is given by
  \be
  H =\frac{1}{3}\left[3H_0 -H_0 \frac{\delta  }{a }+\frac{\dot{\delta } }{a }\right],
  \ee
 where we have denoted
 \be
 \delta  =\sum _{i=1}^3\delta _i=2\delta _1 +\delta _3 .
 \ee

The square of the directional Hubble parameters can be obtained as
\be
H_i^2 \approx H_0^2 -2H_0^2 \frac{\delta _i }{a }+2H_0 \frac{\dot{\delta}_i }{a },\quad i=1,2,3.
 \ee

For the comoving volume $V =\prod  _{i=1}^3a_i$ of the Universe we obtain
 \be
V =a^3 \left[1+\frac{\delta  }{a }\right],
 \ee
and $1/V \approx 1/a^3 $.

To describe the perturbations of the effective energy density and pressure due to the presence of the anisotropic effects induced by the presence of Weyl geometric terms, we introduce the parameter $\beta  $, which allows us to write the perturbations of the effective energy density $\varepsilon$ and pressure $\Psi =\Pi$ of the anisotropic cosmological model in Weyl geometric gravity in the form
\begin{align}\label{P1}
\varepsilon= \rho _{DE}+\rho _{DM}+\beta \left(\rho _{DE}+p_{DE}+\rho _{DM}+p_{DM}\right),
\end{align}
and
\begin{align}\label{P2}
\Psi=\Pi=p_{DE}+p_{DM}+\beta \left(\rho _{DE}+p_{DE}+\rho _{DM}+p_{DM}\right),
\end{align}
respectively, where $\rho_{DE}$ and $p_{DE}$ represent the effective energy density and pressure of the dark energy and dark matter components of the isotropic matter distribution, constructed with the help of the Weyl geometric quantities.  

Therefore, in Eqs.~(\ref{P1}) and (\ref{P2}) we have supposed that the perturbations of the effective thermodynamic quantities of the Weyl geometric cosmological fluid are proportional to the sum of the energy densities and pressures of the effective quantities of the isotropic model.

Hence, in a first order approximation in the metric, the gravitational field equations describing the slightly perturbed isotropic flat FLRW Universe due to Weyl geometric effects  become
\begin{align}
&3\dot{H}_0+3H_0^2 -\left[\dot{H}_0 +H_0^2 \right]\frac{\delta  }{a }+\frac{\ddot{\delta} }{a }\nonumber\\
&=
-\frac{1}{4\kappa^2}\Big[\rho _{DE}+\rho _{DM}+
3\left(p_{DE}+p_{DM}\right) \nonumber\\
&+4\beta  \left(\rho _{DE}+p_{DE}+\rho _{DM}+p_{DM}\right)\Big],
\end{align}
\begin{align}
&\frac{1}{a^3 }\frac{d}{dt}\left[a^3 H_0 \right]+\frac{1}{a^3 }\frac{d}{dt}\left[a^2 H_0 \delta _{+} \right]
+\frac{1}{a^3}\frac{d}{dt}\left[a^2 \dot{\delta }_1 \right]
   \nonumber\\
&-\frac{\delta}{a^4}\frac{d}{dt}\left[a^3 H_0 \right]=\frac{1}{4\kappa^2}(\rho _{DE}-p_{DE}+\rho _{DM}-p_{DM}),
\end{align}
and
\begin{align}
&\frac{1}{a^3 }\frac{d}{dt}\left[a^3 H_0 \right]+\frac{2}{a^3}\frac{d}{dt}\left[a^2 H_0 \delta_{1} \right]-\frac{\delta}{a^4}\frac{d}{dt}\left[a^3 H_0 \right]\nonumber\\
 & +\frac{1}{a^3}\frac{d}{dt}\left[a^2 \dot{\delta}_{3} \right]=
 \frac{1}{4\kappa^2}\left(\rho _{DE}-p_{DE}+\rho _{DM}-p_{DM}\right),\nonumber\\
\end{align}
respectively, where we have denoted $\delta _{+} =\delta _1 +\delta _3$.

The gravitational field equations of the background FLRW geometry are, in Weyl geometric gravity, the generalized Friedmann equations, which give the evolution of the scale factor $a$ as
\be\label{fr1}
3\frac{\dot{a}^2 }{a^2 }=\frac{1}{2\kappa^2}(\rho _{DE}+\rho _{DM}),
\ee
\be\label{fr2}
2\frac{\ddot{a} }{a }+\frac{\dot{a}^2 }{a }=-\frac{1}{2\kappa^2}(p_{DE}+p_{DM}),
\ee
respectively.
Eq.~(\ref{41}), describing the time variation of the rate of the expansion of the volume of the Universe, becomes
\begin{align}
&4\kappa^2\frac{d^2}{dt^2}a^3 +3\left[a^3 \left(p_{DE}-\rho _{DE}+p_{DM}-\rho_{DM}\right)\right]\nonumber\\
&+
4\kappa^2\frac{d^2}{dt^2}(a^2 \delta)
+3\left[a^2 \left(p_{DE}-\rho _{DE}+p_{DM}-\rho_{DM}\right)\right]
\delta=0.
\end{align}
\begin{figure}
	\includegraphics[scale=0.5]{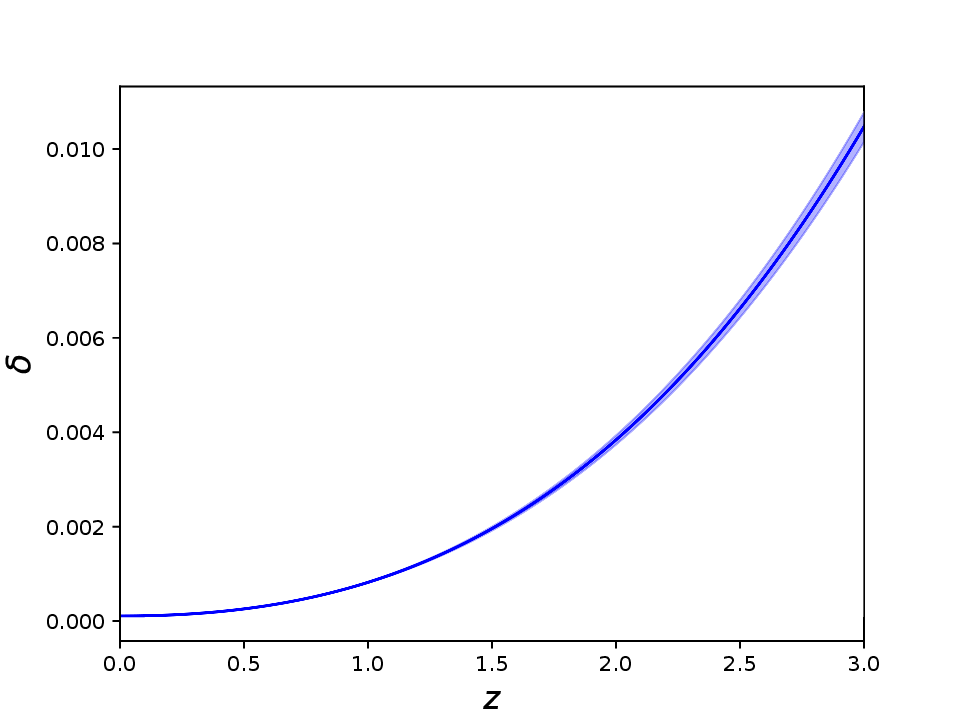}
	\caption{The behavior of the perturbation function $\delta$ as a function of the redshift $z$. The shaded area corresponds to the $1\sigma$ domain for the model parameters \eqref{best}.\label{figdelta}}
\end{figure}
\begin{figure*}
	\includegraphics[scale=0.5]{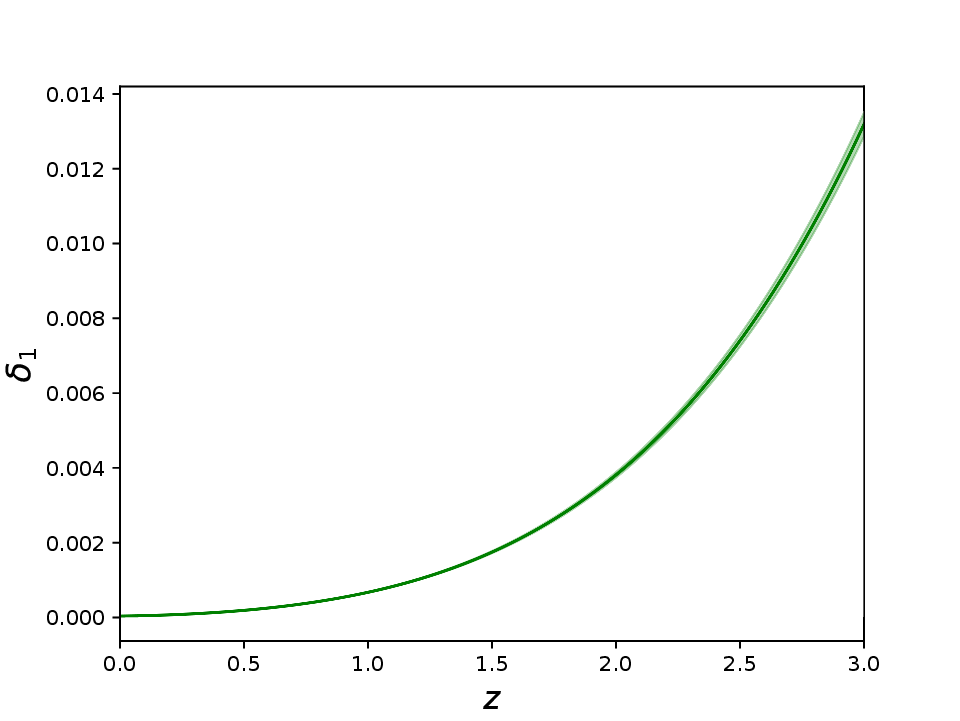}~~\includegraphics[scale=0.5]{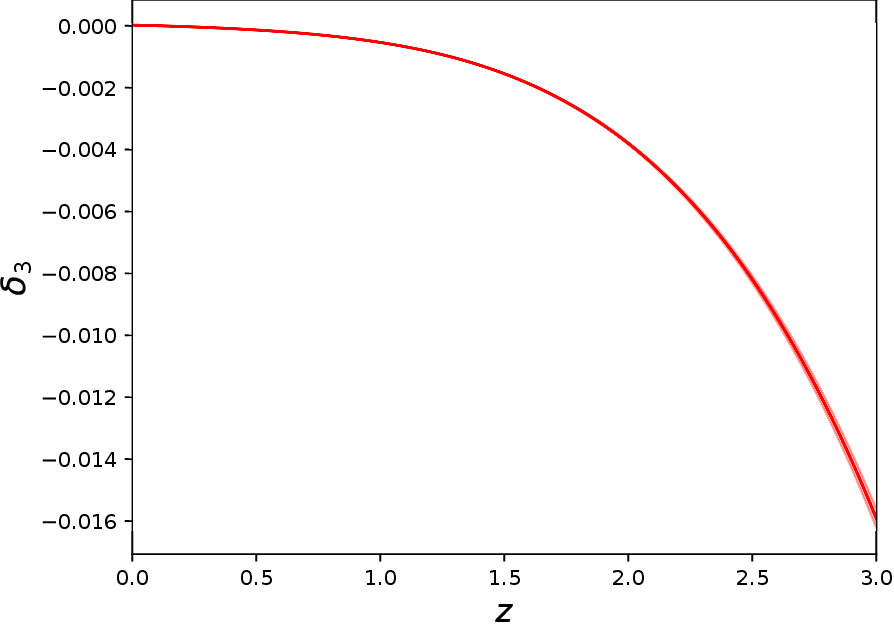}
	\caption{The behavior of the perturbation functions $\delta_1$ (left panel) and $\delta _3$ (right panel) as a function of the redshift $z$. The shaded area corresponds to the $1\sigma$ domain for the model parameters \eqref{best}.\label{figdelta3}}
\end{figure*}
With the use of  Eqs.~(\ref{fr1}) and (\ref{fr2}) we straightforwardly find
\be
\frac{1}{a^3 }\frac{d\left[a^3 H_0 \right]}{dt}=\frac{1}{4\kappa^2}\left(\rho _{DE}-p_{DE}+\rho _{DM}-p_{DM}\right),
 \ee
 \be
 3\dot{H}_0 +3H_0^2 =-\frac{1}{4\kappa^2}\left[\rho _{DE}+\rho _{DM}+3\left(p_{DE}+p_{DM}\right)\right],
  \ee
  and
 \be
  \frac{d^2a^3 }{dt^2}+\frac{3}{4\kappa^2}\left[a^3 \left(p_{DE}-\rho _{DE}+p_{DM}-\rho_{DM}\right)\right]=0,
   \ee
respectively.

Consequently, the gravitational field equations describing the cosmological evolution in the presence of a
small anisotropy in the $z$-direction, induced by the presence of the Weyl geometric effects, are given by
\bea\label{pert1}
\hspace{-0.3cm}\frac{\ddot{\delta} }{a }-\frac{\ddot{a} }{a }\frac{\delta  }{a }=
-\frac{\beta}{\kappa^2} \left(\rho _{DE}+p_{DE}+\rho _{DM}+p_{DM}\right),
\eea
\begin{align}\label{pert2}
\frac{1}{a^3 }\frac{d}{dt}\left[a^2 H_0 \delta _{1} \right]+\frac{1}{a^3}\frac{d}{dt}\left[a^2 \dot{\delta }_1 \right]-\frac{\delta}{a^4}\frac{d}{dt}\left[a^3H_0\right]=0,
\end{align}
\be\label{pert3}
\frac{2}{a^3}\frac{d}{dt}\left[a^2 H_0 \delta_{1} \right]+\frac{1}{a^3}\frac{d}{dt}\left[a^2 \dot{\delta}_{3} \right]-\frac{\delta}{a^4}\frac{d}{dt}\left[a^3H_0\right]=0.
\ee

For a known $H_0$ and $a$,Eqs.~(\ref{pert1})-(\ref{pert3}) represent a system of three ordinary differential equations for the three time dependent functions $\left(\delta _1 ,\delta _3 ,\beta  \right)$, whose solutions indicate the effects of the presence of Weyl geometry on the cosmological evolution.

We consider now the perturbations of the average deceleration parameter $q$ of the anisotropic Bianchi type I Universe, which is defined generally according to
\be
\langle q\rangle=\frac{d}{dt}\frac{1}{H }-1.
\ee
Assuming again that the deviations from isotropy induced by the Weyl geometric effects are small, the average deceleration parameter becomes
\be
\langle q\rangle=q_0+\frac{1}{3}\frac{d}{dt}\left(\frac{1}{aH_0}\left[\delta  -\frac{\dot{\delta } }{H_0}\right]\right),
\ee
where $q_0=d\left[1/H_0 \right]/dt-1=-a \ddot{a} /\dot{a}^2 $ is the deceleration parameter as defined in the FLRW geometry.

By adding Eqs.~(\ref{pert2}) and (\ref{pert3}), we find the second order differential equation describing the evolution of the total perturbation $\delta $ due to the presence of anisotropic effects induced by the Weyl geometry as
\be\label{eqfin}
\ddot{\delta } +4H_0 \dot{\delta } -\left[\dot{H}_0 +5H_0^2 \right]\delta  =0.
\ee
Eq.~(\ref{eqfin}) must be considered together with the two initial conditions
\be
\delta \left(t_0\right)=\delta _0, \qquad \dot{\delta }\left(t_0\right)=H_0\left(t_0\right)\delta _0.
\ee

Hence, we have obtained the result that the time evolution  of the metric perturbations, induced by the presence of Weyl geometric effects, is fully determined by the Hubble parameter of the isotropic FLRW background.
In Fig.~\ref{figdelta} we have plotted the behavior of the perturbation function $\delta$ as a function of redshift, by adopting for $H_0$ the isotropic Weyl geometric form considered in the previous Section. The shaded area corresponds to the $1\sigma$ error in the model parameters given by \eqref{best}.

Once the total perturbation function $\delta  $ is known, the behavior of the deviations function from isotropy along the $x$ and $y$ axes, $\delta_1$, is obtained as a solution of the second order differential equation
\begin{align}
	2\frac{d}{dt}(a^2H_0\delta_1)-2\frac{d}{dt}(a^2\dot{\delta}_1)+\frac{d}{dt}(a^2\dot\delta)-\frac{\delta}{a}\frac{d}{dt}(a^3H_0)=0.
\end{align}
The behavior of $\delta_1$ depends on the evolution of both $a$ and $H_0$. Finally, In Weyl geometric gravity the time evolution of the deviations from isotropy along the $z$ axes can be obtained immediately from the simple relation
\be\label{73}
\delta _3 =\delta  -2\delta _1.
\ee
In Figs.~\ref{figdelta3} we have plotted the behavior of the perturbation functions $\delta_1$ and $\delta_3$ as a function of the redshift.

The parameter $\beta $, describing the perturbations of the effective geometric type density and pressure can be obtained as
\be
\beta  =\kappa^2\frac{(\ddot{H}_0+H_0^2)\delta-\ddot\delta}{a(\rho _{DE}+p_{DE}+\rho _{DM}+p_{DM})}.
\ee

\begin{figure}[h]
	\includegraphics[scale=0.5]{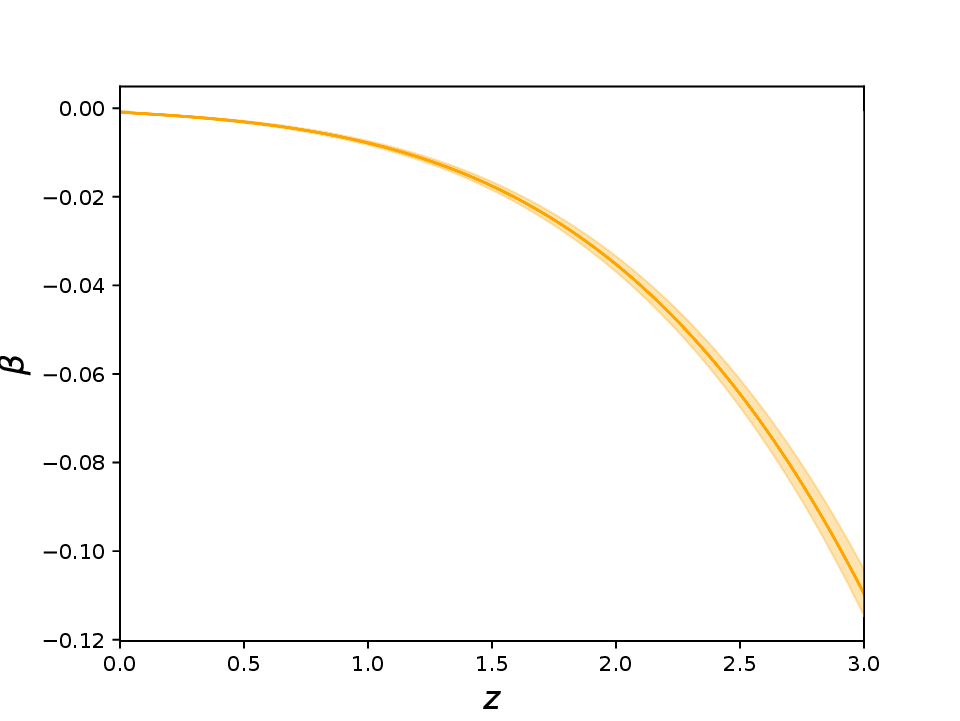}
	\caption{The behavior of the perturbation function $\beta$ as a function of the redshift $z$. The shaded area corresponds to the $1\sigma$ domain for the model parameters \eqref{best}.\label{figbeta}}
\end{figure}

In Figs.~ \ref{figbeta} we have plotted the behavior of the perturbation function $\beta$ as a function of the redshift.

\begin{figure}[h]
	\includegraphics[scale=0.5]{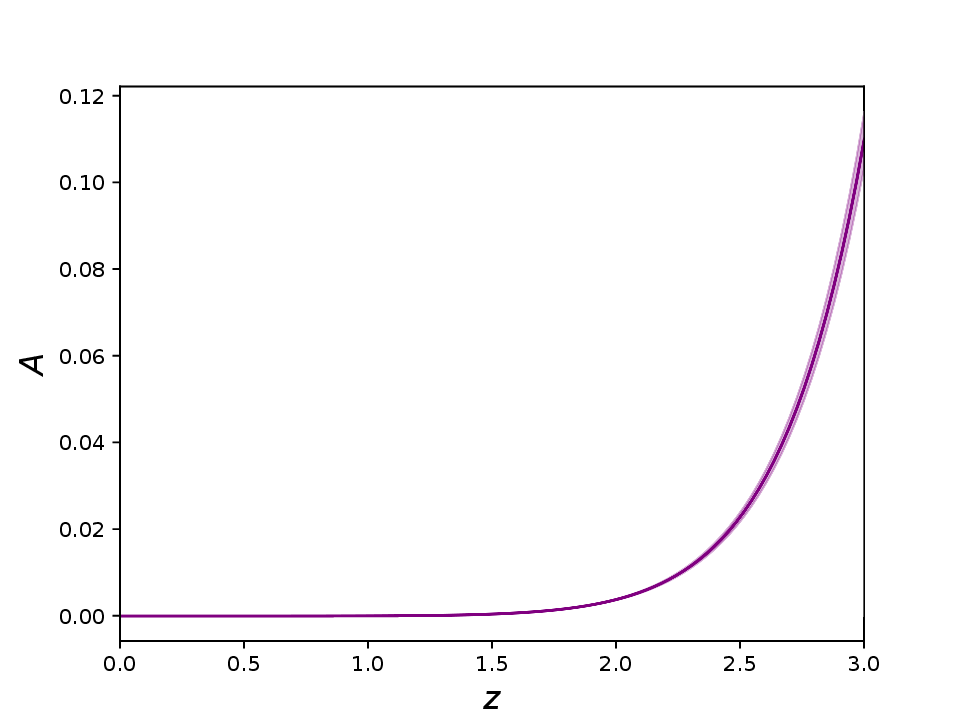}
	\caption{The behavior of the anisotropy parameter $A$ as a function of the redshift $z$. The shaded area corresponds to the $1\sigma$ domain for the model parameters \eqref{best}.\label{Afig}}
\end{figure}

In Fig.~\ref{Afig} we have presented the behavior of the mean anisotropy parameter $A$, in the presence of the Weyl geometric effects, as a function of the redshift $z$.

\begin{figure}[h]
	\includegraphics[scale=0.5]{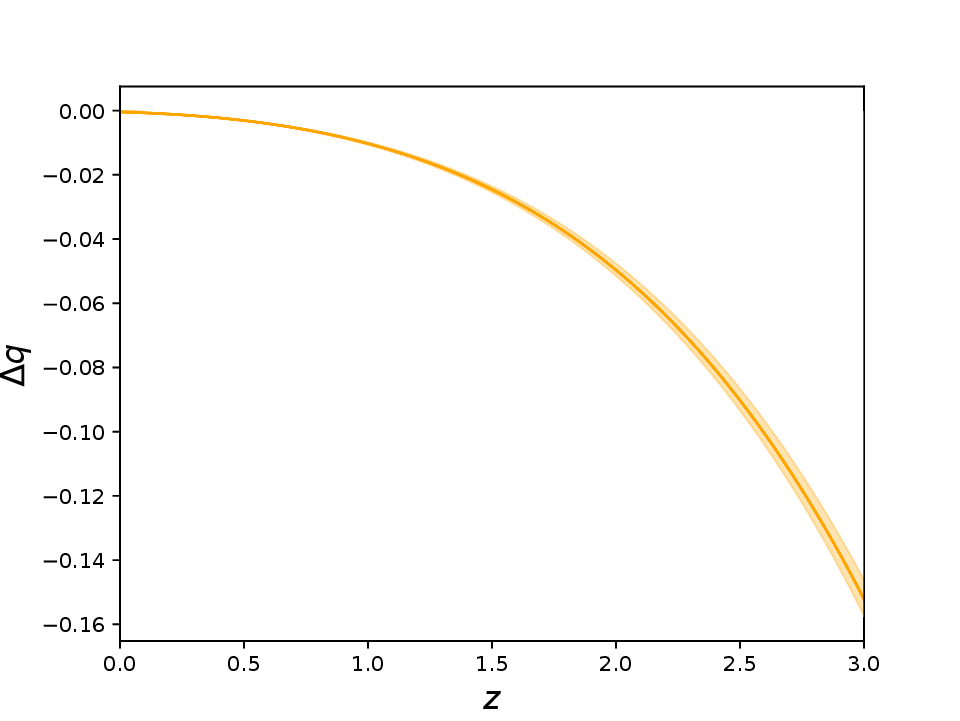}
	\caption{The behavior of the difference between the isotropic and anisotropic deceleration parameter $\Delta q$ as a function of the redshift $z$. The shaded area corresponds to the $1\sigma$ domain for the model parameters \eqref{best}.\label{Dqfig}}
\end{figure}

In Fig.~\ref{Dqfig} we have presented the difference between the isotropic and anisotropic deceleration parameter $\Delta q$, in the presence of the Weyl geometric effects, as a function of the redshift $z$.

\subsection{The quadrupole moment $Q_2$}

The anisotropic expansion of the Universe will also have an influence on photon trajectories, and on the CMBR spectrum. The photons emitted by astrophysical sources located at cosmological distances travel in the Bianchi type I geometry by following the geodesic lines, described by the geodesic equation given by \cite{Tomi1} 
\be
\frac{du^{\mu }}{d\lambda }+\Gamma ^{\mu }_{\alpha \beta}u^{\alpha }u^{\beta }=0,
\ee
where $\lambda $ is the affine parameter along the trajectory. The Christoffel symbols $\Gamma ^{\mu }_{\alpha \beta}$ can be easily obtained from the metric Eq.~(\ref{7}), and they are given by $\Gamma ^{0 }_{i i}=a_i^2H_i$, and $\Gamma ^{ i}_{0 i}=H_i$, $i=1,2,3$ (no summation upon $i$ in the Christoffel symbols).

The four-velocity $u^{\mu }=dx^{\mu }/d\lambda $ of the photons is normalized according to the relation $u^{\mu }u_{\mu }=0$, from which we obtain $\left(u_0\right)^2=a_i^2\left(u^i\right)^2$. Let us now consider the emission of two photons, occuring at the times  $t_0 = t_e$ and  $t_1 = t_e + \delta \tau$, respectively, where $\delta \tau \ll t_e$. in the first order of approximation in $\delta \tau $, after taking the difference of the two normalization conditions for photons,  we obtain \cite{Tomi1}
\be\label{86}
u^0\frac{d }{d\lambda }\delta \tau (\lambda)=\sum_{i=1}^3{a_i\dot{a}_i\left(u^i\right)^2\delta \tau (\lambda )}+O\left(\delta \tau ^2\right),
\ee
where $\lambda$ is the wave length of the radiation. By denoting by $\delta \tau  \left(\lambda _r\right)$ the time difference between the received signals, and by introducing the redshift $z$, defined according to $1+z\left(\lambda _e\right)=\delta \tau \left(\lambda _r\right)/\delta \tau \left(\lambda _e\right)$, with the use of Eq.~(\ref{86}) we  obtain the equation
\be
\frac{d}{d\lambda }\ln(1+z)=\frac{1}{u^0}\sum_{i=1}^3{a_i\dot{a}_i\left(u^i\right)^2}.
\ee

The components $u^i$, $i=1,2,3$, of the photon velocity follow immediately from the geodesic equation of motion, and they are given by
\be\label{sg}
\frac{du^i}{d\lambda }+2\frac{\dot{a}_i}{a_i}u^iu^0=0, \qquad  i=1,2,3,
\ee
giving
\be
u^i(t)=\frac{u_{0i}}{a_i^2(t)}, \qquad  i=1,2,3,
\ee
where $u_{0i}$, $i=1,2,3$ are arbitrary constants of integration, which can be determined by taking into account that the present day values of the scale factors are normalized according to $a_i\left(t_0\right) = 1$. 

 We can reparameterize the affine parameter $\lambda $ without any impact on the physical behavior of the radiation, and thus we normalize the present day photon four-velocities $u^i\left(t_0\right) = \hat{u}^i$, $i=1,2,3$, according to $\sum_{i=1}^3\left(\hat{u}^i\right)^2= 1$. Moreover, we describe the unit vector $\hat{u}$ with the help of the angles
$\left(\hat{u}_x, \hat{u}_y, \hat{u}_z\right) = \left(\sin \theta \cos \phi, \sin \theta \sin \phi, \cos \theta\right)$, which give the arrival angles of the photon beams to the observer, as estimated at the present time. After the substitution of the photon velocities into the redshift definition, we find the relation \cite{Tomi1}
\be
1+z\left(\hat{\vec{u}}\right)=\sqrt{\sum_{i=1}^3{\frac{\hat{u}_i^2}{a_i^2}}},
\ee
or, equivalently,
\be\label{ecc}
1+z\left(\hat{\vec{u}}\right)=\frac{1}{a_1}\sqrt{1+\hat{u}_y^2e_y^2+\hat{u}_z^2e_z^2}.
\ee

In Eq.~(\ref{ecc} we have introduced the eccentricities $e^2_y$ and $e_z^2$,  defined according to
\be
e_y^2=\left(\frac{a_1}{a_2}\right)^2-1, \qquad  e_z^2=\left(\frac{a_1}{a_3}\right)^2-1.
\ee
Since in the anisotropic Weyl geometric gravity model we have assumed  $a_1=a_2$, we immediately obtain $e_y^2=0$, and
\be
e_z^2=\frac{2}{a}\left[\delta _1(t)-\delta _3(t)\right],
\ee
respectively.

The multipole spectrum $Q_l$ of the CMBR is described in terms of the coefficients in the spherical expansion of the temperature anisotropy field. The observationally important quadrupole term $Q_2$ is obtained as \cite{Tomi1}
\bea
Q_2&=&\frac{2}{5\sqrt{3}}\sqrt{e_z^4+e_y^4-e_z^2e_y^2}=\frac{2}{5\sqrt{3}}e_z^2 \nonumber\\
  &=&\frac{4}{5\sqrt{3}}\frac{1}{a}\left[\delta _1(t)-\delta _3(t)\right].
\eea

In the case of the anisotropic Weyl geometric gravity cosmological models  one obtains
\begin{align}
	Q_2=\frac{2}{5\sqrt{3}}(1+z)(5\delta_1-2\delta),
\end{align}

\begin{figure}[h]
	\includegraphics[scale=0.5]{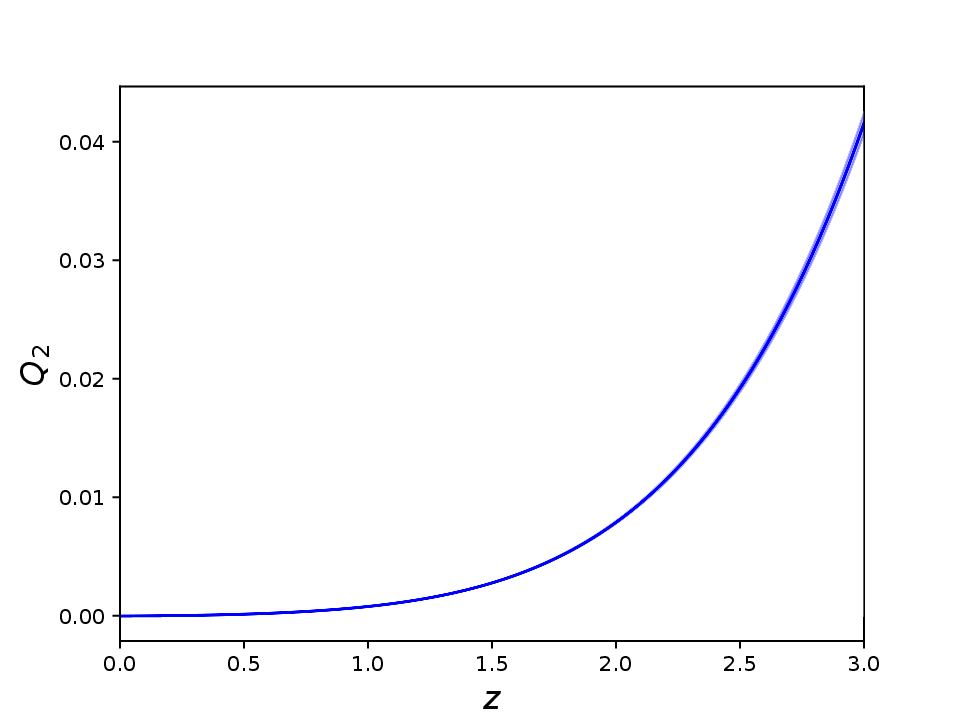}
	\caption{The behavior of the quadrupole term  $Q_2$ as a function of the redshift $z$ in the anisotropic Weyl geometric gravity cosmological models. The shaded area corresponds to the $1\sigma$ domain for the model parameter \eqref{best}.\label{Q2fig}}
\end{figure}

In Fig.~ \ref{Q2fig} we have plotted the quantity $Q_2$ as a function of the redshift $z$.

The quadrupole moment measured by the Planck satellite is $Q\equiv \left(\Delta T/T\right)_Q\sim 4.5\times 10^{-6}$ \cite{No}. Hence, observational results on the CMBR could be used to constrain the numerical values of the model parameters, and the predictions of the Weyl geometric gravity cosmological models.

The temperature distribution of the CMBR can be obtained from the relation
\be
T\left(\hat{\vec{u}}\right) =\frac{T_{*}}{1+z\left(\hat{\vec{u}}\right)},
\ee
where by  $T_{*}$ we have denoted the last scattering temperature, which is independent on the direction. But in the presence of anisotropies in the metric,  photons
travelling  from distinct directions will be redshifted differently. The spatial average $\bar{T}$ of the temperature field can be computed from the relation $4\pi \bar{T}=\int{T\left(\hat{\vec{u}}\right)d\Omega _{\hat{\vec{u}}}}$. The anisotropies in the temperature field can be obtained as
\bea
\hspace{-0.9cm}\delta T\left(\hat{\vec{u}}\right)&=&1-\frac{T\left(\hat{\vec{u}}\right)}{\bar{T}} \nonumber\\
\hspace{-0.9cm}    &=&1-\frac{T_{*}}{\bar{T}}\frac{1}{\sqrt{1+2\left[\delta _1\left(t_0\right)-\delta _3\left(t_0\right)\right]\cos ^2\theta}}.
\eea

Hence, a detailed comparison of the predictions of the anistropic Weyl geometric cosmological model with the CMBR data could lead to obtaining significant constraints on the presence of the anisotropy in the Universe, and of the Weyl geometric effects.   

\section{Discussions and final remarks}\label{sect4}

In the present paper, we have investigated the theoretical possibility that our Universe may be modeled in terms of a Weylian geometric structure, in which the gravitational properties are described by the metric tensor, and two other fields, namely, a scalar and a vector field. The considered theory implements strictly the idea of the conformal invariance of the gravitational interaction. The problem of the role of the conformal symmetry, or of the Weyl gauge group in gravitational theories has attracted a lot of interest recently.  Gerard ’t Hooft advanced the idea that conformal symmetry is an exact symmetry of the natural laws, which is spontaneously broken during the evolution of the Universe \cite{G1}. Therefore,  conformal symmetry may be as important for physics as the Lorentz invariance of natural laws is. Moreover, the breaking of the conformal symmetry may allow us to understand in geometric terms the small scale structure of gravity, and to obtain new insights in the physics of
the Planck scale. An approach to gravity based on the assumption that conformal symmetry is an exact local, but spontaneously broken symmetry, was investigated in \cite{G2}. 

Conformal symmetry also plays a fundamental role in the Conformal Cyclic Cosmology (CCC) model \cite{P1,P2,P3}, in which the Universe is assumed to exist as o set of eons, geometric structures corresponding to time oriented spacetimes. Eons have interesting mathematical properties, and, for example, they possess, as a direct result of their conformal compactification,  spacelike null infinities.

In the Weyl geometric theory considered in this work the idea of conformal invariance is rigorously implemented at the level of both geometry and matter. The starting point is the conformally invariant geometric quadratic Weyl action, with gravitational Lagrangian $\tilde{R}^2$, supplemented by the square of the strength of the Weyl vector field, and by a matter term.  A new perspective on the theory can be obtained by introducing an auxiliary scalar field in the formalism, thus transforming the initial vector-tensor theory into a scalar-vector tensor theory, linear in the Weyl (and Ricci) scalars. The physical implications of this theory have been investigated in detail elsewhere \cite{Gh10,Gh11, Gh12,Gh13}, and it was shown that it has many attractive features, including the possibility of representing a bridge between gravity and elementary particle physics. We have analyzed in detail the cosmological implications of the theory, by assuming a particular form, additive in the baryonic matter Lagrangian, and the Weyl vector, of the effective, conformally invariant matter Lagrangian. By using this simplified model we have considered in detail two general classes of cosmological models.

The first class we have considered is represented by flat, homogenous and isotropic FLRW type models. The symmetry of the problem imposes a fixed structure of the Weyl vector, which has only a non-zero temporal component. After obtaining the generalized Friedmann equations, we have investigated the vacuum model, and a baryonic matter filled Universe. In both cases we have compared the predictions of the Weyl geometric gravity theory with a small set of observational data for the Hubble parameter, as well as with the predictions of the $\Lambda$CDM model.

The vacuum Weyl geometric gravity model gives a relatively good description of the observational data for the Hubble, indicating that the geometric contributions from the Weyl geometry can successfully simulate dark energy and dark matter, and, to some extent, even baryonic matter. The statistical analysis predicts a value of the Hubble constant which is closer to the SHOES value \cite{Ca3} than to the Planck value \cite{Ca1}. The deceleration parameter of the vacuum model is relatively consistent with the $\Lambda$CDM predictions in the redshift range $0<z<1$, but significant differences do appear at higher redshifts. The exponent $n$, giving the contribution of the square of the Weyl vector to the effective, conformally invariant matter energy momentum tensor, is almost equal to 2, and hence $\mathcal{L}_m=L_m+\beta \left(-\omega ^2\right)^2$. The temporal component of the Weyl vector takes negative values, and it is a decreasing function of the redshift (an increasing function of time). On the other hand, the scalar field $\Phi$ is a monotonically increasing function of the redshift, increasing rapidly with increasing $z$.

Adding baryonic matter to the Weyl geometric FLRW model does not change drastically the parameters of the cosmological evolution. In the presence of matter the model also gives an acceptable description of the observational data, as well as of the $\Lambda$CDM model, but without succeeding in reproducing it exactly. Similar differences do appear in the case of the deceleration parameter, with the Weyl geometric gravity model indicating much higher values of $q$ at higher redshifts. Significant differences do appear in the predictions of the behavior of the matter energy density, with the Weyl geometric gravity model predicting much higher matter density values as compared to the $\Lambda$CDM model. In the presence of the baryonic matter the temporal component of the Weyl vector is a monotonically decreasing function of the redshift, taking negative values, while the scalar field is a positive, monotonically increasing function of the redshift. 

There is compelling observational evidence \cite{Ca1} that the Universe is isotropic, and homogeneous on large scales. However, the Planck Collaboration experiments have  not provided yet a conclusive observational proof for the cosmological isotropy.  Moreover, recently a number of observations have questioned the nature of the geometry of cosmological spacetime itself, and suggested the existence of deviations from the homogeneous and isotropic FLRW geometry. In this respect we may mention the quadrupole-octupole alignment problem, the lack of correlations on large angular scales, and
the hemispherical power asymmetry \cite{N1, N2}. All these observations seem to suggest a violation of statistical isotropy in the Universe, and of the scale-invariance of the primordial spectrum fluctuations. Hence, the investigation of the anisotropic cosmological models may not be of purely theoretical interest, but could lead to the explanation of some observed features of the cosmological expansion.  Hence, the above observations, as well as the Planck results suggest that the possibility of the existence of a large scale cosmological anisotropy in the Universe cannot be neglected {\it a priori} in the theoretical models.  

In the present work we have also considered an analysis of the simplest anisotropic, Bianchi type I cosmological models, in the Weyl geometric gravity theory. . We have assumed that the Weyl vector has two nonzero components, $\omega _\mu=\left(\omega _0,0,0,\omega _3\right)$. As a first step in our analysis we have reformulated the cosmological evolution equations in terms of an effective energy density, and of two effective pressures, which combine the effects of the baryonic matter and of the Weyl geometry.  We have also considered that the deviations from isotropy are very small, and that the anisotropic properties of the Universe can be described as a small perturbation  of the homogeneous and isotropic FLRW background metric. For the case of the Bianchi type I geometry considered in the present paper we have explicitly derived  the perturbation equations, by assuming that the scale factors of the Bianchi type I Universe are given by $a_i=a+\delta a_i$, $i=1,2,3$, with $\delta a_i$ satisfying the condition $\delta _i \ll a$, $i=1,2,3$. If this condition is not satisfied, non-linear effects must also be taken into account, and included in the analysis of the cosmological evolution.

Therefore, due to the presence of the $z$ component of the Weyl vector,  and of the anisotropic effective pressure distribution,  the Weyl type Universe would achieve some anisotropic properties, and its geometry will slightly differ from the standard FLRW one. Our perturbative analysis has shown that the behavior of the total perturbations of the FLRW metric $\delta$, given by Eq.~\ref{eqfin}, depend only, in the considered linear approximation, on the Hubble function of the isotropic model. Once this function is known, by numerically integrating Eq.~(\ref{eqfin}) one can obtain a full description of the behavior of the anisotropic perturbations in the cosmology of the Weyl geometric gravity theory.

The total perturbation of the metric, shown in Fig.~\ref{figdelta}, is a monotonically increasing positive function of the redshift, or a monotonically decreasing function of the cosmological time. $\delta$ is relatively constant in the redshift range $0<z<1$, and increases rapidly afterwards, making at higher redshifts the Universe more and more anisotropic. $\delta_1$, represented in Fig.~\ref{figdelta3}, has a similar behavior as $\delta$, but $\delta _3$, shown in the same Figure, takes negative values, and decreases with the redshift, thus increasing the $z$ axis anisotropy with increasing $z$. 

An important parameter describing the cosmological properties of the anisotropic Weyl geometric gravity cosmological models  is the parameter $\beta$, defined in Eqs.~(\ref{P1}) and (\ref{P2}), respectively, and which describes the perturbations of the effective densities and pressure of the model, which include the contributions of the Weyl vector, and of the scalar field. The variation of $\beta$, represented in Fig.~\ref{figbeta}, indicates that for the considered background cosmological model this quantity takes negative values, and it decreases with the redshift.   

The anisotropy parameter $A$, depicted in Fig.~\ref{Afig}, is practically zero in the redshift range $0<z<2$, and it increases rapidly for $z>2$. This indicates a rapid increase, and the strong presence, of the anisotropies at higher redshifts. Hence, in the present cosmological scenario, the initially anisotropic Universe isotropizes in the large time limit, with the present day Universe being isotropic on average. The difference of the isotropic and anisotropic deceleration parameters, represented in Fig.~\ref{Dqfig}, shows an increase of the difference at higher redshifts, the differences between the two quantities varying slowly in the redshift range $0<z<1$.     

In our analysis we have also pointed out the possibilities of observationally testing the Weyl geometric gravity cosmological model, which could be done, for example, by using the Planck data \cite{Ca1}. We have explicitly obtained the expression of the quadrupole $Q_2$ for an anisotropic Universe as functions of the deviations $\delta _1$ and $\delta _3$ from the isotropic FLRW geometry. The variation of $Q_2$ as a function of redshift is presented in Fig.~\ref{Q2fig}). $Q_2$ is practically constant in the redshift range $0<z<1$, but it rapidly increases at higher redshifts.  One can use the obtained expression of $Q_2$ to obtain some constraints on the parameters of the anisotropic Weyl geometric cosmological model. We have already obtained constraints on the isotropic model parameters from the study of the luminosity distance by using the observational data coming from the  type Ia supernovae.

In the present paper we have investigated, in the framework of Weyl geometric gravity,  both homogeneous isotropic and anisotropic cosmological models. We have also performed a full comparison of the predictions of the isotropic cosmological model with the observations, and we have found that the model can give a satisfactory description of the observational data up to a redshift of $z=2$. We have also assumed the existence on large cosmological scales of small deviations from isotropy, and we have obtained  a full description of the behavior of the Bianchi type I models in Weyl geometric gravity. In this study we have obtained the basic theoretical tools necessary for the in depth investigation of the implications of the Weyl geometric effects in cosmology.  The impact of the Weyl geometry on the cosmological evolution, the  observational implications of the presence of this geometry, as well the possibilities of its observational testing, will be considered in  future studies.

\section*{Acknowledgments}

The work of T.H. is supported by a grant of the Romanian Ministry of Education and Research, CNCS-UEFISCDI, project number PN-III-P4-ID-PCE-2020-2255 (PNCDI III).

\end{document}